\pdfoutput=1
\documentclass[11pt]{article}
\usepackage[final]{acl}

\usepackage{times}
\usepackage{latexsym}
\usepackage{amsmath,amssymb,amsfonts}
\usepackage{graphicx}
\usepackage{booktabs}
\usepackage{longtable}
\usepackage{caption}
\usepackage{stfloats}
\usepackage{multirow} 
\usepackage{comment}
\usepackage{mfirstuc,textcase,xspace,graphicx}

\usepackage{longtable}
\usepackage{wrapfig}
\usepackage{comment}  
\usepackage{algorithmic}
\usepackage{float}
\usepackage{xcolor,colortbl} 
\usepackage{amsmath,amsfonts} %
\usepackage{graphicx}%
\usepackage{textcomp}%
\usepackage{array}%
\usepackage{pifont}%
\usepackage{amsthm}%
\usepackage{bm}
\usepackage[linesnumbered,ruled,vlined]{algorithm2e}%
\usepackage{setspace}%
\usepackage{multirow}%
\usepackage{siunitx}
\usepackage{footnote}%
\usepackage[protrusion=true,expansion=true,activate={true,nocompatibility},final,tracking=true,kerning=true,spacing=true,stretch=10,shrink=10,disable=footnote]{microtype}
\usepackage{blindtext}
\usepackage{ctable} 
\usepackage{pifont}
\usepackage{booktabs}
\usepackage{makecell}
\usepackage{tabularx}
\usepackage{enumitem}
\graphicspath{ {image/} }

\usepackage{titlesec}

\usepackage{microtype}
\usepackage{inconsolata}
\usepackage{url}
\usepackage{pifont}
\newcommand{\cmark}{\ding{51}}
\newcommand{\xmark}{\ding{55}}

\DeclareUnicodeCharacter{2009}{,}
\DeclareUnicodeCharacter{2248}{\ensuremath{\approx}}

\newcommand{\mycolor}[1]{\textcolor{black}{#1}}

\title{CIS-BWE: Chaos-Informed Speech Bandwidth Extension}

\author{Tarikul Islam Tamiti$^1$, Tonmoy Das$^2$, Nursadul Mamun$^3$, Anomadarshi Barua$^4$ \\
$^{2,3}$Chittagong University of Engineering and Technology, Bangladesh \\
$^{1,4}$George Mason University, USA\\
}

\date{}

\begin{document}
\maketitle

\begin{abstract}


We design CIS-BWE, a novel adversarial Bandwidth Extension (BWE) framework that introduces two chaos-informed discriminators - Multi-Resolution Lyapunov Discriminator (MRLD) and  Multi-Scale Detrended Fractal Analysis Discriminator (MSDFA) - for capturing the deterministic chaos from speech. MRLD exploits Lyapunov exponents to capture nonlinear chaotic fluctuations. MSDFA exploits detrended fluctuation analysis to quantify fractal-like, long-range temporal chaotic correlations. To the best of our knowledge, MRLD and MSDFA are included here for the first time with a complex-valued adversarial network to explore the chaotic study of speech reconstruction. We also introduce a novel complex-valued and
dual-stream generator, which uses our newly proposed ConformerNeXt as a core block with Lattice interactions, acting as a gating mechanism by enabling controlled mixing of information across streams. We extensively optimize our design across five resolutions and use depth-wise separable convolutions to make our model lightweight yet powerful. Our CIS-BWE is tested with a de facto English and French dataset for clean and noisy speech for generalization. It achieves better performance across a total of nine subjective and objective evaluation metrics with a 40x reduction in discriminator size and overall 0.5x fewer parameters, establishing a new baseline in the BWE task.

\end{abstract}

\vspace{-01.0em}
\section{Introduction and Related Work}
\vspace{-0.2em}
\label{sec:intro}

Bandwidth Extension (BWE) can enhance speech by reconstructing missing high frequencies from low-frequency data \cite{li2015deep}. It is critical for wide range of Natural Language Processing (NLP) applications, such as Text-To-Speech (TTS) and Automatic Speech Recognition (ASR) \cite{haws2019cyclegan}. 
\mycolor{BWE preserves critical spectral cues and thus improves recognition accuracy on low-bandwidth inputs for TTS and ASR tasks.} 



\mycolor{
Early BWE models  mainly focused on magnitude spectrogram enhancement \cite{li2015deep, sui2024tramba, abreu2024aeromamba, hu2022accear,lu2025explicit}, neglecting the phase due to its noisy patterns and relying on vocoders for audio reconstruction \cite{ho2025investigation, liu2022neural, tamiti2025practical}. However,  \cite{gerkmann2012phase,lu2025explicit,tamiti2025high,tamiti2025practical} show that enhancing} phase together with magnitude (real and imaginary) yields higher perceptual  audio \cite{yin2020phasen} at the cost of increased computational complexity. However, none of them considers the \textit{chaotic modeling} of speech generation and hence, they miss the opportunity to improve their performance with \textit{less complexity}. We refer to Appendix \ref{subsec:Chaotic_PropertiesofSpeechGeneration} to understand the presence of chaos in  speech.


Speech production is fundamentally non-linear with the presence of deterministic chaos due to the complex interaction between airflow and deformable vocal tract \cite{herzel1994analysis}. These chaotic dynamics are different from the ones present in phase and sensitively dependent on 
slow-variant features \cite{michael1999speech}. Although the Multi Period Discriminator (MPD) and Multi Scale Discriminator \cite{kong2020hifi}, Multi Band Discriminator \cite{yang2021multi}, and Multi-Resolution Spectrogram Discriminator \cite{jang2021univnet} are proposed to determine the nonlinear cues among temporal structures, they lack a chaotic feature extraction framework. Therefore, they fail to capture the intricate chaotic features, leading to residual artifacts and temporal blurring  \cite{kim2021fre}. Moreover, they are usually parameter-heavy, resulting in extra computational overhead.

In this paper, we propose a novel class of chaos-informed discriminators for capturing the deterministic chaos, which State-of-the-Art (SOTA) work overlooks. We design two chaos-inspired  discriminators - Multi-Resolution Lyapunov Discriminator
(MRLD) and Multi-Scale Detrended Fluctuation Analysis Discriminator (MSDFA). MRLD uses Lyapunov exponents \cite{oseledec1968multiplicative} to capture rapid and nonlinear chaotic fluctuations. MSDFA uses detrended fluctuation analysis \cite{peng1994mosaic} to capture fractal-like temporal correlations. This is the first time that MRLD and MSDFA have been proposed to be included with complex-valued Generative Adversarial Networks (GANs) to explore the chaotic study of audio reconstruction, to the best of our knowledge. Our extensive design optimization across five resolutions and the use of depth-wise separable convolution make MRLD and MSDFA lightweight yet powerful, which requires 0.5x fewer parameters, a 40x reduction in discriminator size compared to SOTA \cite{lu2024towards}, with better performance across a wide range of subjective and objective metrics. 


We also introduce a novel complex-valued and dual-stream generator, which uses ConformerNeXt as a core block with Lattice interactions, acting as a gating mechanism by enabling controlled mixing of information across streams. Proposed ConformerNeXt is a combination of Transformer-based Conformer \cite{gulati2020conformer} for capturing global context and Convolution Neural Network (CNN)-based ConvNeXt \cite{liu2022convnet} for capturing local context efficiently present in speech. This optimized generator architecture simultaneously enhances magnitude and phase stream within a compact yet efficient architecture. We name our proposed model \textbf{CIS-BWE} (\textbf{C}haos-\textbf{I}nformed \textbf{S}peech BWE). Our key technical contributions are: 

\textbf{(1)} For the first time, we introduce chaos-informed and parameter-efficient non-linear discriminators to capture deterministic chaos.

\textbf{(2)} We introduce dual-stream ConformerNeXt with Lattice interactions for controlled feature mixing for high-fidelity speech reconstruction.

\textbf{(3)} We extensively evaluate performance using \mycolor{seven} objective metrics: LSD, PESQ, STOI, SI-SDR, SI-SNR, NISQA-MOS, \mycolor{and WER}; two subjective metrics: MOS and Pairwise preference test; three computation metrics: MAC, FLOP, and RTF (see Sections \ref{subsec:EvaluationMetrics}, \ref{subsec:Computational Complexity}, and \ref{subsec:Subjective Test} for full form).

\begin{figure*}[htbp]
  \centering
\includegraphics[width=0.85\textwidth,height=0.18\textheight]{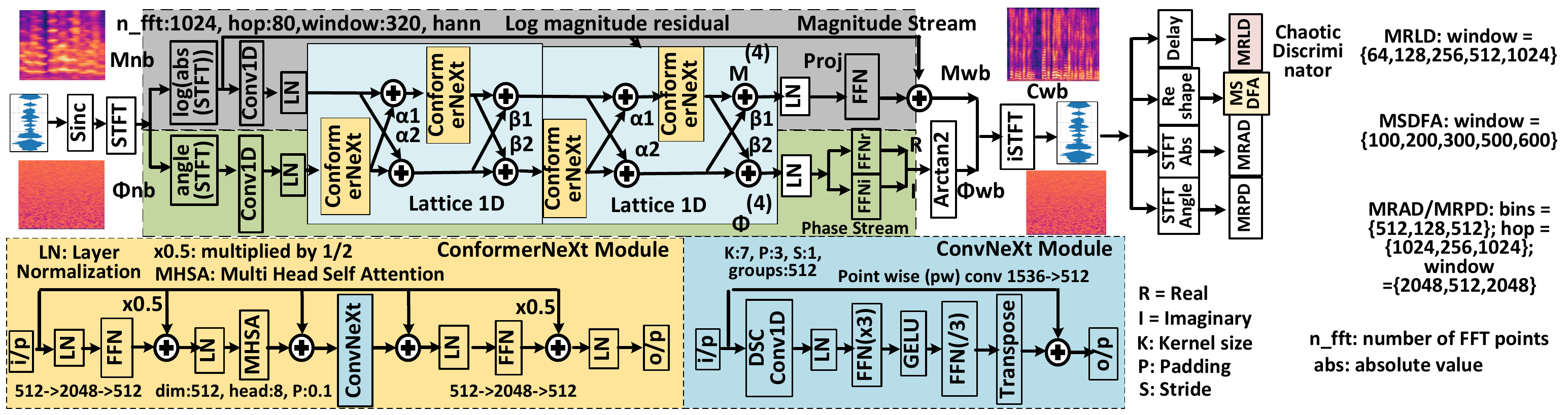}
\vspace{-0.750em}
  \caption{Our proposed CIS-BWE, containing Lattice net, ConformerNeXt, and chaos-informed discriminators.}
  \label{fig:overall_architecture}
  \vspace{-01.60em}
\end{figure*}

\vspace{-0.43em}
\section{Overall Architecture}
\label{sec:overallarchitecture}
\vspace{-0.3em}

The proposed architecture is illustrated in Fig. \ref{fig:overall_architecture}.

\vspace{-0.43em}
\subsection{Generator Architecture}
\label{sec:genarchitecture}
\vspace{-0.3em}


Due to the necessity of phase along with amplitude \cite{yin2020phasen, lu2024multi}, the generator of our CIS-BWE parallely receives both the amplitude and phase cues from the two computed synchronized feature maps. The magnitude spectrogram is obtained by taking the logarithm of the absolute value of the Short-Time Fourier Transform (STFT), and the phase spectrogram is obtained by taking the angle of the STFT. We define the narrow-band magnitude and phase spectrograms by $\mathbf{M}_{\mathrm{nb}}$ and $\boldsymbol{\Phi}_{\mathrm{nb}}$, respectively, in Eqn. \ref{eqn:magnitudeandphase}.


\vspace{-0.43em}
{\small
\begin{equation}
\begin{aligned}
\mathbf{M}_{\mathrm{nb}} \in \mathbb{R}^{B \times F \times T}, 
\quad
\boldsymbol{\Phi}_{\mathrm{nb}} \in \mathbb{R}^{B \times F \times T}
\label{eqn:magnitudeandphase}
\end{aligned}
\vspace{-0.5em}
\end{equation}
}
\vspace{-1.32em}

where $B$ is the batch size, $F$ is the number of frequency bins ($F = \tfrac{n\_{\mathrm{FFT}}}{2} + 1$), 
and $T$ is the number of time frames. The generator's forward pass consists of the following three main stages:

\textbf{a) Dual Stream Processing:} The generator initially processes the narrow-band magnitude and phase in two separate streams and merges them into a common latent space at the end, harmonizing heterogeneous inputs into a unified projection. 


\textbf{b) Lattice Block Interaction:} The Lattice block \cite{Luo2020LatticeNet} ensures continuous controlled mixing of magnitude and phase streams that enables explicit exchange of information, reweighing each stream's contribution, and faster convergence. Although the model could get faster inference for the absence of this mixing, however, the absence of the mixing results in error accumulation in the streams,``muffled" artifacts in the reconstructed wideband audio signal, and unstable training. Therefore, we apply two successive one-dimensional Lattice blocks (\textbf{Lattice1D}), each of which interleaves the magnitude and phase streams via criss-cross connections by learnable scalars, denoted by $\alpha_1, \alpha_2, \beta_1$, and $\beta_2$ in Fig. \ref{fig:overall_architecture}. These scalars perform as a gating function by controlling the strength of cross-stream injection. These scalars are trained end-to-end and dynamically learn ``where" and ``how-much" cross-stream interactions are required. 


Within each Lattice block, we insert a new module named \textbf{ConformerNeXt}. We design ConformerNeXt by replacing the standard Conformer’s \cite{gulati2020conformer} convolutional sub-module with a ConvNeXt \cite{liu2022convnet} block. This replacement significantly enhances the model's capacity by combining ConvNeXt's powerful hierarchical spatial feature extraction with Conformer's global self-attention. Our proposed ConformerNeXt takes two inputs instead of one, as opposed to \cite{Luo2020LatticeNet}. Moreover, the pre- and post-processing convolutions used in the original Lattice Nets in \cite{Luo2020LatticeNet}  are omitted. This variation provides the best choice for the inter-stream connectivity.

As there are two Lattice blocks and each contains one ConformerNeXt per stream, the full generator employs a total of four ConformerNeXt with cross‐stream residual connections. After trying various combinations of connections and the number of ConformerNeXt, we find that this design provides the best trade-off between performance and a reduced number of parameters. The granular-level implementation details, like the number of filter channels, attention headcounts for ConformerNeXt are shown in Fig. \ref{fig:overall_architecture} (see Appendix \ref{subsec:Parameter Breakdown for Generators}).

\textbf{c) Residual Prediction and Synthesis:}  After the final Lattice stage, separate output heads estimate the wideband residuals $\mathbf{M}^{(4)}$ and $\mathbf{\Phi}^{(4)}$, which are then passed through Layer Normalization (LN) and Linear Projection (Proj) blocks implemented by the Feed Forward Neural Network (FFN) to estimate the wide band residuals. 
The magnitude branch outputs a log-magnitude residual that is added to the narrow-band input to isolate and recover only missing high-frequency information, rather than remodeling the entire spectrum. 

Moreover, previous work \cite{yin2020phasen} shows that due to the noisy nature of the phase, it is very difficult to estimate the phase directly. 
To overcome these difficulties in direct phase estimation, the phase stream's output is fed into two FFNs for predicting ``pseudo-real (R)'' and ``pseudo-imaginary (I)'' residuals, shown in Eqn. \ref{eqn:phase_residual}. Finally, the wide-band phase is recovered by the ``arctan2'' function by stacking the magnitude and phase branch, and using Inverse STFT, the wide-band audio is reconstructed (see Eqn. \ref{eqn:istft}).

\vspace{-01.2em}
{\small
\begin{equation}
\begin{aligned}
  R = \mathrm{FFN}_r\bigl(\mathrm{LN}(\mathbf{\Phi}^{(4)})\bigr) ;
  I = \mathrm{FFN}_i\bigl(\mathrm{LN}(\mathbf{\Phi}^{(4)})\bigr); \\
   \Phi_{\mathrm{wb}} = \mathrm{atan2}(I, R)
\end{aligned}
\label{eqn:phase_residual}
\end{equation}
}
\vspace{-02.1em}

{\small
\begin{equation}
\begin{aligned}
\mathbf{C}_{\mathrm{wb}}= e^{\mathbf{M}_{\mathrm{wb}}} \bigl(\cos\Phi_{\mathrm{wb}} + i\sin\Phi_{\mathrm{wb}}\bigr)
\end{aligned}
\label{eqn:istft}
\end{equation}
\vspace{-0.9em}
}


\vspace{-0.9em}
\subsection{Chaos-Informed Nonlinear Discriminator}
\label{subsec:chaosinormeddiscriminator}

Speech production is fundamentally a \emph{non-linear dynamical process characterized by deterministic chaos} ~\cite{little2007exploiting}. Discriminators used in traditional GANs ~\cite{tian2020tfgan}, ~\cite{donahue2018adversarial}, ~\cite{kumar2019melgan} typically minimize the distance between reconstructed and original speech based on raw waveforms or spectrogram slices, but fail to detect those nonlinear chaotic cues. Therefore, generators produce over-smoothed and dull spectra ~\cite{cao2024vnet}. Here, we design two chaos-inspired nonlinear discriminators - MRLD and MSDFA. 
This is the first time that these two discriminators have been proposed to be included with complex-valued GANs to explore the chaotic study of audio reconstruction. They analyze long-range and hidden formant trajectories and micro-transients across equally spaced windows, output chaos-aware feature maps, and penalize any mismatch in sub-harmonic richness. Our approach results in a 40x reduction in discriminator size, 0.5x fewer parameters, and more realistic acoustics (see Sections \ref{subsec:generator_ablation}, \ref{subsec:Discriminator_Ablation_Study}, \& \ref{subsec:ComparativeAnalysiswithBaselines}) with less over-smoothed spectra compared to SOTA models. 

\textbf{a) Multi‐Resolution Lyapunov Discriminator (MRLD):} We introduce Lyapunov Exponents (LE) \cite{oseledec1968multiplicative, wolf1985determining} to capture the rapid, nonlinear fluctuations and sensitivity to initial conditions in speech that spectrogram-based losses overlook. The LE is a measure of nonlinear dynamics used to quantify the rate of separation of infinitesimally close trajectories. Therefore, MRLD penalizes the mismatches in the Lyapunov spectra of real and generated signals and drives the generator to reproduce authentic deterministic chaotic behavior, yielding more lifelike speech.

\begin{figure}[ht!]
  \centering
\includegraphics[width=0.48\textwidth,height=0.13\textheight]{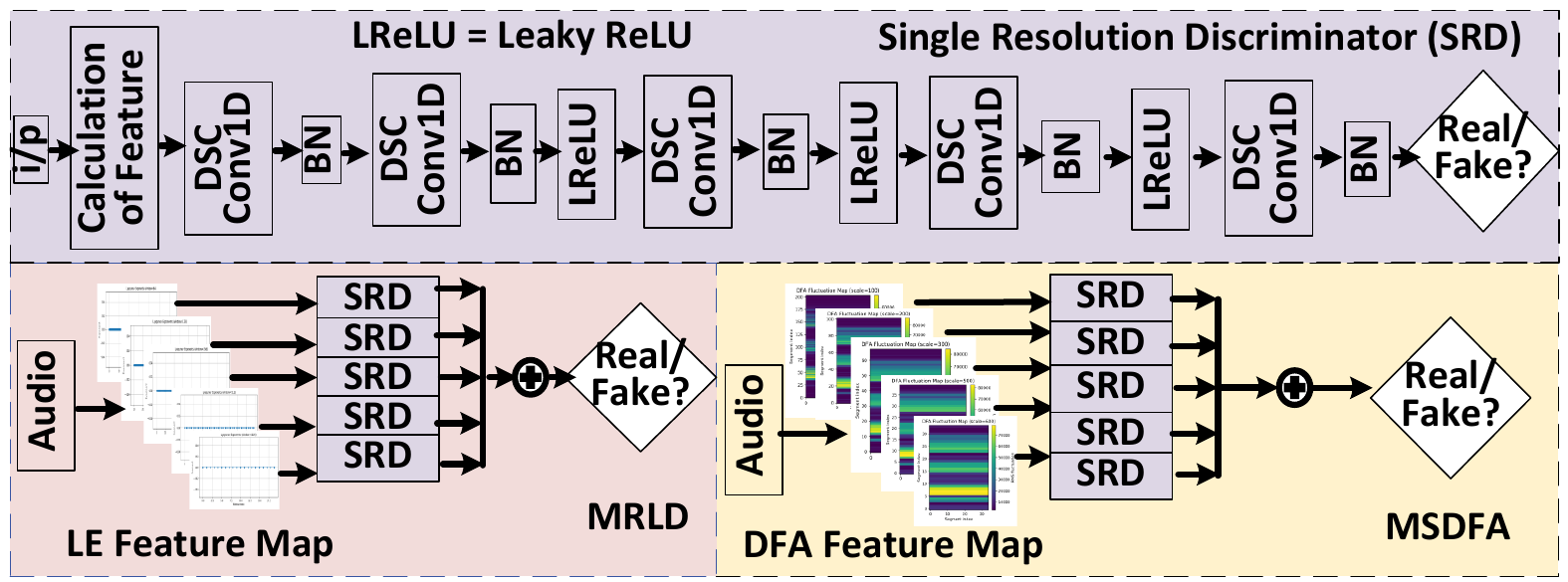} 
 \vspace{-01.905em}
  \caption{Implementation details of the MRLD, MSDFA, and SRD. We refer to Appendix \ref{subsec:Parameter Breakdown for Discriminators} for details.}
  \label{fig:discriminator}
  \vspace{-0.515em}
\end{figure}

A pseudo-code \ref{alg:MRLD} is provided to explain how MRLD is implemented. MRLD divides each waveform into five non-overlapping windows $w\in\{64,128,256,512,1024\}$, computes local LE via delay-embedding and nearest-neighbor divergence, and maps each segment to a single divergence rate (lines 1-7). Then, MRLD feeds these five exponent maps into five separate Single Resolution Discriminator (SRD). The detailed structure of the SRD is shown in Fig. \ref{fig:discriminator}. SRD uses a five-layer depthwise-separable (DSC) 2D Convolution with kernel size 5 (stride 2 for the first four layers, final kernel 3, 235k parameters) to discriminate real versus generated dynamics (refer to Appendix \ref{subsec:Parameter Breakdown for Discriminators} for details).

\vspace{-0.8em}
\begin{algorithm}
\small
\caption{Pseudo-code of MRLD}
\begin{algorithmic}[1]
\label{alg:MRLD}
\REQUIRE Raw waveform $x$, window sizes $\mathcal{W} = \{64, 128, 256, 512, 1024\}$
\ENSURE Predicted label $y \in \{\text{real}, \text{generated}\}$

\STATE $\mathcal{F} \gets []$ \COMMENT{Initialize feature list}

\FORALL{$w \in \mathcal{W}$}
    \FORALL{segment $x_i^w$ in $x$ with window size $w$}
        \STATE Delay-embed $x_i^w$ into vectors $\{y_j\}$ using dimension $d$, delay $\tau$
        \STATE Compute $\lambda_i^w = \text{Avg}_j \left[ \frac{1}{\Delta} \log \left( \frac{\|y_{j+\Delta} - y_{j'+\Delta}\| + \epsilon}{\|y_j - y_j'\| + \epsilon} \right) \right]$
        \STATE Append $\lambda_i^w$ to $\mathcal{F}$
    \ENDFOR
\ENDFOR

\STATE Normalize and reshape $\mathcal{F}$ for SRD input
\STATE $y \gets \text{SRD}_{\text{MRLD}}(\mathcal{F})$
\RETURN $y$
\end{algorithmic}
\end{algorithm}
\vspace{-01.17em}

\vspace{-0.605em}

\textbf{b) Multi‐Scale Detrended Fluctuation Analysis Discriminator (MSDFA):}  We introduce Detrended Fluctuation Analysis (DFA) \cite{peng1994mosaic} to quantify fractal-like, long-range temporal correlations that conventional spectrogram losses overlook. Therefore, by computing how root-mean-square fluctuations $F(n)$ grow with window size, MSDFA supervises the generator to ensure natural-sounding dynamics across syllabic, phonemic, and sub-phonemic scales. If omitted, the adversarial framework leads to muffled prosody even when amplitude and phase discriminators are present.

A pseudo-code \ref{alg:MSDFA} is provided to explain how MSDFA is implemented. We integrate classical DFA at five $n\!\in\!\{100,200,300,500,600\}$ samples, tile each $F(n)$ into a fixed $S\times S$ map, and feed the resulting tensor into five separate SRD modules. The SRD (see Fig. \ref{fig:discriminator} and Appendix \ref{subsec:Parameter Breakdown for Discriminators}) has a five-layer DSC 2D CNN (BN + LeakyReLU), whose adversarial loss back-propagates through the tiling, making MSDFA a light (≈247K parameters) yet powerful plug-in discriminator for BWE problems. 

\begin{table*}[ht]
  \centering
  \scriptsize
  \setlength{\tabcolsep}{2pt}
  \renewcommand{\arraystretch}{0.8}
  \begin{tabular}{m{2.3cm} | m{4.5cm}| m{8.8cm}}
    \toprule
      \textbf{Loss function} 
      &\textbf{Equation \,(MSE = Mean Square Error)}
      & \textbf{Terms }  \\
    \midrule

    Magnitude Loss  
    & $\mathcal{L}_{\text{mag}} = \lambda_{\text{mag}} \cdot \mathrm{MSE}(\hat{M}, M)$ 
    & $\!M,\hat M$: ground‐truth and generated STFT magnitudes. $\lambda_{\text{mag}}$ = 45 \\
\midrule
    Phase Loss 
    & $\mathcal{L}_{\text{pha}} = \lambda_{\text{pha}} \cdot (\mathcal{L}_{\text{IP}} + \mathcal{L}_{\text{GD}} + \mathcal{L}_{\text{IAF}})$
    & IP = Instantaneous Phase difference, GD = Group Delay difference, IAF = Instantaneous Amplitude-Frequency difference. $\lambda_{\text{pha}}$ = 100.  \\
\midrule
    Complex STFT Loss 
    & $\mathcal{L}_{\text{com}} = \lambda_{\text{com}} \cdot \mathrm{MSE}(\hat{C}, C)$
    & \(C\) and \(\hat{C}\) are the complex-valued STFTs of the target and predicted signals. $\lambda_{\text{com}}$ = 90. \\
\midrule
    Self-Consistency Loss 
    & $\mathcal{L}_{\text{stft}} = \lambda_{\text{stft}} \cdot \mathrm{MSE}(\hat{C}, \tilde{C})$
    & \(\tilde{C}\) =  STFT of the waveform reconstructed from predicted magnitude and phase. $\lambda_{\text{stft}}$ = 90. \\
\midrule
   Feature Matching Loss
   & $\mathcal{L}_{\text{fm}} = \sum_{d \in \mathcal{D}} \lambda_{d} \cdot \mathrm{MSE}(f_d^{\text{real}}, f_d^{\text{fake}})$
   &  Feature maps from discriminator \(d\), \(d\in\{\mathrm{MRLD,\,MSDFA,\,MRAD,\,MRPD}\}\) \\
\midrule
  Generator Hinge Loss 
  & $\mathcal{L}_{\text{adv}} =\sum_{d\in\mathcal{D}}\lambda_{d}\cdot\mathbb{E}_{\hat{x}\sim p_G}\left[max(0,1-D_d(\hat{x}))\right]$
  &  \(\hat x\sim P_G\): generated samples;
  \(D_d(\hat x)\): discriminator \(d\) score;
  \(\lambda_d\): weight on \(d\)’s adversarial term    \\
\midrule
   Discriminator Hinge loss 
  & $\mathcal{L}_D^{(d)} =
\mathbb{E}_{x \sim p_{\text{data}}} \left[ \max(0, 1 - D_d(x)) \right]  +
\mathbb{E}_{\hat{x} \sim p_G} \left[ \max(0, 1 + D_d(\hat{x})) \right]$  
   & \(x\sim P_{\rm data}\): real samples;
\(\hat x\sim P_G\): generated samples;
\(\max(0,1 - D_d(x))\): real‐hinge term (\(D_d(x)\ge1\));
\(\max(0,1 + D_d(\hat x))\): fake‐hinge term (\(D_d(\hat x)\le-1\))\\

    \bottomrule
  \end{tabular}
\vspace{-1.3em}
    \caption{Generator and discriminator loss functions. Code will be released after the acceptance of the paper.}
      \label{tab:lossfunction}
      \vspace{-2.78em}
\end{table*}

\textbf{c) Multi‐Resolution Amplitude \& Phase Discriminators (MRAD \& MRPD):}  In addition to MRLD and MSDFA, we also use MRAD and MRPD in our adversarial framework. MRAD ensures that amplitude transients are captured in different granularities. MRPD stabilizes group delay and explores harmonic-phase relationships. We refer to \cite{lu2024multi} for the implementation details of MRAD and MRPD. Similar to \cite{lu2024multi}, we use three resolutions, such as frequency bins = [512,128,512], hop sizes = [1024,256,1024], and window lengths = [2048,512,2048]. Each resolution is fed to a 5‐layer 2D CNN (varied kernels/strides, weight‐norm) (see Appendix \ref{subsec:Parameter Breakdown for Discriminators}). 

\vspace{-01.0em}
\begin{algorithm}[ht!]
\small
\caption{Pseudo-code of MSDFA}
\label{alg:MSDFA}
\begin{algorithmic}[1]
\REQUIRE Input waveform $x(t)$, scales $\mathcal{N} = \{100, 200, 300, 500, 600\}$
\ENSURE Real/fake score $y \in \mathbb{R}$

\STATE $\mathcal{F} \gets []$
\FORALL{$n \in \mathcal{N}$}
    \STATE Compute DFA fluctuation $F(n)$ on $x(t)$
    \STATE Tile $F(n)$ into fixed-size map $M^{(n)} \in \mathbb{R}^{S \times S}$ and Append $M^{(n)}$ to $\mathcal{F}$
\ENDFOR

\STATE $T \gets \text{stack}(\mathcal{F}) \in \mathbb{R}^{S \times S \times 5}$
\STATE $y \gets \text{SRD}_{\text{MSDFA}}(T)$ 
\RETURN $y$
\end{algorithmic}
\end{algorithm}
\vspace{-01.07em}

\vspace{-0.4em}
\subsection{Loss Functions}
\vspace{-0.215em}



\textbf{Generator Losses:} We propose a total of six different generator loss functions shown in Table \ref{tab:lossfunction}. Magnitude loss encourages accurate spectral amplitude reconstruction. Phase loss ensures faithful temporal alignment and phase continuity. Complex STFT loss jointly enforces faithful amplitude and phase reconstruction. Self consistency loss enforces synthesis consistency. Feature matching loss is critical as it penalizes subtle nuances enforced by non-linear, amplitude, and phase discriminators' feedback. 
Adversarial loss encourages realism in the  waveform generated across multiple perceptual dimensions. The total Generator Loss is shown as:

\vspace{-01.43em}
{\small
\begin{equation}
\mathcal{L}_G = \mathcal{L}_{\text{mag}} + \mathcal{L}_{\text{pha}} + 
\mathcal{L}_{\text{com}} + \mathcal{L}_{\text{stft}} + 
\mathcal{L}_{\text{fm}} + \mathcal{L}_{\text{adv}}.
\end{equation}
}
\vspace{-1.86em}

\textbf{Discriminator Losses:} 
Each discriminator \(D_d\) is trained using a hinge loss objective, which specializes discriminators to become powerful critics of unnatural patterns by matching with the perceptual distribution of real speech (see Table \ref{tab:lossfunction}). The total discriminator loss is shown in Eqn. \ref{eqn:discriminitorloss}.


\vspace{-1.0em}
{
\scriptsize
\begin{equation}
\mathcal{L}_D = \sum_r \mathcal{L}_D^{\text{MRLD}}  + 
\sum_s \mathcal{L}_D^{\text{MSDFA}}  + 
\sum_r \mathcal{L}_D^{\text{MRAD}}   + 
\sum_r \mathcal{L}_D^{\text{MRPD}}   
\label{eqn:discriminitorloss}
\end{equation}
}
\vspace{-1.0em}

where $\mathcal{L}_D^{\text{MRLD}}$, $\mathcal{L}_D^{\text{MSDFA}}$, $\mathcal{L}_D^{\text{MRAD}}$, and $\mathcal{L}_D^{\text{MRPD}}$ are MRLD, MSDFA, MRAD, and MRPD losses, respectively, for each resolution/scales. 

\vspace{-0.65em}
\section{Comprehensive Analysis}
\label{sec:Comprehensive_Result_Analysis}
\vspace{-0.55em}

\subsection{Evaluation Metrics}
\label{subsec:EvaluationMetrics}
\vspace{-0.3em}

We assess 
quality of reconstructed speech using \mycolor{seven} metrics: Log‐Spectral Distance (LSD) \cite{115972} to quantify fine‐grained spectral deviations; Short‐Time Objective Intelligibility (STOI) \cite{taal2011algorithm} to evaluate speech intelligibility; Perceptual Evaluation of Speech Quality (PESQ) \cite{rix2001perceptual} to predict quality in line with human judgments; Scale‐Invariant SDR (SI‐SDR) \cite{le2019sdr} as a general distortion metric invariant to amplitude scaling; Scale‐Invariant SNR (SI‐SNR) \cite{luo2018tasnet} to specifically gauge noise‐related distortion; Non‐Intrusive Speech Quality Assessment (NISQA‐MOS) \cite{mittag2021nisqa} for reference‐free estimation of perceptual speech quality; \mycolor{and Word Error Rate (WER) for ASR evaluation.} 

\vspace{-0.64em}
\subsection{Hyperparameter and Configuration} 
\label{subsec:Hyperparameter_Configuration}
\vspace{-0.2em}


The training of CIS-BWE involves carefully chosen hyperparameters. Learning rate is initialized with $2\times10^{-4}$ with exponential decay after each epoch with a decay factor of 0.999. AdamW optimizer with $\beta_1$ = 0.8, and $\beta_2$ = 0.99, and a weight decay of 0.01 are used for stable convergence. We use a batch size of 16 to balance between computational efficiency and memory utilization. All models are trained for a total of 50 epochs and per epoch takes around 25 minutes. \textbf{We list all the hyperparameters in Appendix \ref{subsec:hyperparameters}.}  
We use NVIDIA RTX-4090 GPU and AMD Ryzen 9 7950X3D 16-Core CPU for computation. 

\vspace{-0.64em}
\subsection{Dataset and Preprocessing} 
\label{subsec:Dataset_Preprocessing}
\vspace{-0.2em}

We use the English VCTK (v0.92) ~\cite{yamagishi2019cstr} (110 speakers) and MLS French corpus \cite{Pratap2020MLSAL} (114 speakers), to test in multi-lingual and cross-speaker setup. There are in total $\sim$1045 hours of speech in the two datasets with a sampling rate of 16 and 48 kHz. We load 16/48 kHz files, convert to mono channel, remove silence parts, downsample to simulate band-limited audio, sinc interpolate, and align length-wise. All test data have unseen speakers that are not included in the train data (i.e., cross speaker). 
We refer to Appendix \ref{sec:dataset_preprocessing}, \ref{sec:dataset_dataloader}, \ref{sec:feature_extraction} for detailed explanation. 

\vspace{-0.7em}
\subsection{Discriminator Ablation Study} 
\label{subsec:Discriminator_Ablation_Study}
\vspace{-0.2em}

Table~\ref{table:discriminator_ablation_study} represents the ablation results reported based on five objective evaluation metrics. 

\textbf{Row\textcircled{ \scriptsize 1}:} When we only use MRPD (phase) and MSDFA (fractal dynamics), the model receives feedback only on fine-grained periodicity and long-range temporal self-similarity, resulting in the poorest performance (NISQA-MOS = 2.32). However, a higher SI-SNR indicates that the generator can generate phase-consistent results without generating natural-sounding envelopes.

\textbf{Row\textcircled{\scriptsize 2} and \textcircled{\scriptsize 3}:} MRLD (deterministic chaos) and MSDFA show an improved performance (NISQA-MOS = 3.58) as MRLD encourages realistic chaos, but without amplitude cues, the performance is lower. Using MRAD and MRPD in row \textcircled{\scriptsize 3}, we obtain NISQA-MOS = 4.07 because we have both magnitude and phase cues. MRAD enforces the correct amplitude distributions, and MRPD aligns the notorious phase relationships. These combinations also score significantly better in LSD (1.11) and PESQ (1.61). These results show us the efficacy of the MRAD and MRPD and make them ``indispensable" in our design choice.

\textbf{Row\textcircled{\scriptsize 4} and \textcircled{\scriptsize 5}:} We use MRLD instead of MRPD along with MRAD in row \textcircled{\scriptsize 4}. This performs on par with MRAD and MRPD, as both phase and Lyapunov exponents capture two different types of deterministic chaos. Once again, when we use both MRLD and MRPD along with MRAD, we get an increase in NISQA-MOS to 4.14 from 4.08, which confirms the necessity of MRLD in capturing deterministic chaos. 

\vspace{-0.0em}
\begin{table}[ht!]
  \scriptsize
  \setlength{\tabcolsep}{1.3pt}  
\begin{tabular}{l c c c c c c c c c c}
\toprule
SL & MPD & MRAD & MRPD & MRLD  & MSDFA 
    & LSD    & STOI  & PESQ    & SNR & N-MOS \\
\hline

 1 & \xmark & \xmark & \cmark & \xmark & \cmark 
    & 1.22 & 0.89 & 2.05   &  9.47  & 2.32 \\

 2 & \xmark & \xmark & \xmark & \cmark  & \cmark 
    & 1.20 & 0.85 & 1.55   & 7.47  & 3.58 \\

 3 & \xmark & \cmark & \cmark & \xmark  & \xmark 
    & 1.11 & 0.86 & 1.61 &   7.87  & 4.07 \\

4 &  \xmark & \cmark & \xmark & \cmark  & \xmark  
    & 1.09 & 0.86 & 1.65   & 8.42  & 4.08 \\


 5& \xmark & \cmark & \cmark & \cmark  & \xmark  
    & 1.06 & 0.85 & 1.58  & 7.78  & 4.14 \\
\rowcolor{blue!30}
 6 & \xmark & \cmark & \cmark & \cmark  & \cmark 
    & 1.10 & 0.87 & 1.66  & 8.11 & 4.29\\
\hline 
\multicolumn{11}{c}{} \\[-4.5pt]
\multicolumn{11}{c}{Evaluating MPD with our proposed discriminators}\\ [2pt]
\hline
 7 & \cmark & \xmark & \xmark & \cmark & \xmark 
    & 1.23 & 0.85 & 1.53   & 6.95  & 3.58 \\
 8 & \cmark & \xmark & \xmark & \cmark  & \cmark 
    & 1.04 & 0.85 & 1.55   & 7.15  & 3.65 \\
9 & \cmark & \xmark & \xmark & \xmark  & \cmark 
    & 1.24 & 0.85 & 1.52   & 7.08  & 3.79 \\
\rowcolor{gray!30}
10 &  \cmark & \cmark & \cmark & \xmark & \xmark  
    & 1.11 & 0.85 & 1.56   & 6.68  & 4.18 \\
\hline 
\multicolumn{11}{c}{} \\[-4.5pt]
\multicolumn{11}{c}{Comparison of parameters among MPD and our proposed discriminators}\\ [2pt]
\hline 
\multicolumn{11}{c}{} \\[-4.5pt]
11 &  {\color{magenta}22M} &  600.2k & 600.2k & {\color{blue}235.5k} & {\color{blue}247.7k} & & &  &  &   \\
\bottomrule
\end{tabular}
\vspace{-0.8765em}
\caption{Ablation study on discriminators with their sizes for 2→16 kHz range. Here, N-MOS = NISQA-MOS and SNR = SI-SNR. Our MRLD + MSDFA in row \textcircled{ \scriptsize 6} has in total 40x smaller parameters (22M vs 483.2k) compared to MPD in row \textcircled{ \scriptsize 10} with better performance.}
\label{table:discriminator_ablation_study}
\vspace{-01.5em}
\end{table}

\textbf{Row\textcircled{\scriptsize 6}:} 
We give fractal analysis feedback to the generator by MSDFA along with MRLD, MRPD, and MRAD. These combined features provide strong cues to the generator, which results in significant boosts to NISQA-MOS from 4.14 to 4.29. Therefore, this set of combinations is used in our proposed CIS-BWE architecture.


\vspace{-0.45em}
\subsection{Comparison with MPD} 
\label{subsec:Comparison with MPD}
\vspace{-0.2em}

As SOTA models \cite{lu2024multi,lu2024towards} use MPD, we compare the performance of our proposed MRLD + MSDFA with MPD in row \textcircled{\scriptsize 7} to \textcircled{\scriptsize 10} of Table \ref{table:discriminator_ablation_study}.

\textbf{Row\textcircled{\scriptsize 10}:} The combination of MPD + MRPD + MRAD gives NISQA-MOS of 4.18, which is lower than our proposed combinations of MRLD + MSDFA + MRAD + MRPD in Row\textcircled{\scriptsize 6}. This statement also holds for other metrics as well. As MRAD + MRPD is common in both cases, this shows that the MRLD + MSDFA performs better than MPD alone. Moreover, we are getting better performance with only 483.2k parameters in total for MRLD + MSDFA, compared to 22M parameters of MPD. This is a significant finding as our MRLD + MSDFA gives better performance compared to MPD with 40x smaller parameters (22M vs 483.2k). This will provide a stepping stone for smaller models in edge devices without sacrificing performance (refer to Appendix \ref{subsec:Parameter Breakdown for Discriminators} for details on discriminator).

\begin{table*}[ht]
  \centering
  \scriptsize
  \setlength{\tabcolsep}{1.7pt}
  \renewcommand{\arraystretch}{0.9}
  \begin{tabular}{@{}l l l c c c c c c c c @{}}
    \toprule
    \textbf{SL}
      & \textbf{Architecture} 
      &\textbf{Size}
      & \textbf{Discriminators} 
      & \textbf{H}
      & \textbf{LSD} $\downarrow$ 
      & \textbf{STOI} $\uparrow$ 
      & \textbf{PESQ} $\uparrow$ 
      & \textbf{SI-SDR} $\uparrow$ 
      & \textbf{SI-SNR} $\uparrow$ 
      & \textbf{NISQA-MOS} $\uparrow$ \\
    \midrule
    1 & ConvNeXt$_{16}$ & 106.34M & [MRLD + MSDFA + MRA/PD]x3 & - &1.12 & 0.86 & 1.63 & 7.89 & 7.88 & 4.31 \\
    2 & ConformerNeXt$_{16}$  & 223.2M & [MRLD + MSDFA + MRA/PD]x3 & 8 &1.10 & 0.87 & 1.70 & 7.93 & 7.90 & 4.44 \\
    3 & ConformerNeXt$_{4}$, Linear & 64.23M & [MRLD + MSDFA + MRA/PD]x3 & 8 & 1.12  &0.86 & 1.57 & 7.66 & 7.62 & 3.72 \\
    4 & ConformerNeXt$_{4}$ & 32.7M & [MRLD + MSDFA + MRA/PD]x3 & 8 & 1.09 &  0.87 & 1.71 & 8.56 & 8.54 & 4.03 \\
    5 & ConformerNeXt$_{4}$ & 33.5M & [DSC(MRLD+MSDFA) + MRA/PD]×5 & 4 & 1.10 & 0.87 & 1.67 & 8.24 & 8.20 & 4.25 \\
    6 & ConformerNeXt$_{4}$, MLP /4 & 16.67M & [DSC(MRLD+MSDFA+MRA/PD)]×5 & 4 &1.12 & 0.87 & 1.68 & 8.01 & 7.98 & 3.60 \\
    \rowcolor{blue!30}
    7 & ConformerNeXt$_{4}$ (Proposed) &33.5M & [DSC(MRLD + MSDFA)]x5 + [MRA/PD]x3 & 8 & 1.10 & 0.87 & 1.66 & 8.14 & 8.11 & 4.29 \\
    \bottomrule
  \end{tabular}
\vspace{-1em}
    \caption{Generator architecture ablation study for 2→16 kHz range. SI-SNR and SI-SDR use dB unit. H = number of attention heads in the multi-head self attention of the ConformerNeXt block.} 
      \label{tab:generator_ablation}
      \vspace{-2.58em}
\end{table*}

\vspace{-0.4em}
\subsection{Generator Architecture Ablation Study}
\label{subsec:generator_ablation}
\vspace{-0.2em}


To determine the optimal generator configuration, we performed three systematic ablations on core block selection, inter-stream connectivity, network depth, and Multilayer Perceptron (MLP) expansion ratio, summarized in Table~\ref{tab:generator_ablation} for 2→16 kHz range.

\textbf{Core Block Selection:} We consider two blocks: ConvNeXt and ConformerNeXt. The reason for choosing ConvNeXt is that we want to demonstrate the better performance of our ConformerNeXt over the SOTA ConvNext \cite{lu2024multi,lu2024towards}. We separately use a total of 16 ConvNext (denoted by ConvNeXt$_{16}$ in row \textcircled{\scriptsize 1}) and ConformerNeXt (denoted by  ConformerNeXt$_{16}$ in row \textcircled{\scriptsize 2}). The row \textcircled{\scriptsize 2} achieves the highest NISQA-MOS of 4.44, a +0.13 gain over ConvNeXt in row \textcircled{\scriptsize 1}, along with improvements in LSD (1.10 vs 1.12), STOI (0.87 vs 0.86), and PESQ (1.70 vs 1.63). In this way,  we find the supremacy of the ConformerNeXt over ConvNeXt as a core block.

\textbf{Inter-Stream Connectivity:} We compare linear and Lattice Net for coupling magnitude and phase streams. Lattice Net in row \textcircled{\scriptsize 4}  consistently outperforms the linear stream in row \textcircled{\scriptsize 3}  in terms of all  \mycolor{seven} metrics. Therefore, we use Lattice Net as a cross-stream interaction for its superior controlled mixing of amplitude and phase stream via learnable scalars' gating mechanism (see Section \ref{sec:genarchitecture}). 


\textbf{Depth and Head Count:} After getting the best core block and cross-connection scheme, we optimize the number of ConformerNeXt and head count in the generator. Reducing from 16 to 4 ConformerNeXt blocks (row \textcircled{\scriptsize 2} vs row \textcircled{\scriptsize 4}) yields a compact model with a 7x reduction in size with compromising a small performance but still better/similar to SOTA models \cite{lu2024multi, lu2024towards} in Table \ref{tab:Comparative_analysis}. We further evaluated multi-head self-attention by reducing from 8 to 4 heads (row \textcircled{\scriptsize 5}), leading to a minor drop in NISQA-MOS (4.25 vs 4.29) compared to row \textcircled{\scriptsize 12}. We also test the hidden dimension of the linear network with one-fourth (row \textcircled{\scriptsize 6}), which degrades NISQA-MOS to 3.60. 

\vspace{-0.7em}
\subsection{Making the Discriminator Efficient}
\label{subsec:Discriminator_Efficient}
\vspace{-0.2em}


The performance of our generator is highly correlated with the number of scales or resolutions used in the discriminators. In rows \textcircled{\scriptsize 1} to \textcircled{\scriptsize 4} of Table ~\ref{tab:generator_ablation}, we use three windows or scales to calculate features from three different resolutions. However, when we increase the number of scales to 5, the feature maps capture more fine-grained as well as coarse-grained temporal patterns to provide better performance, which is shown in rows \textcircled{\scriptsize 5} to \textcircled{\scriptsize 7} of Table ~\ref{tab:generator_ablation}. Due to the larger scale of 5, though it might take slightly more train time due to calculate more features, our discriminators can guide the generator for more faithful reconstruction without any increase in the number of parameters.

Moreover, we use DSC in our discriminators to shrink their sizes. In rows \textcircled{\scriptsize 1} to \textcircled{\scriptsize 4} of Table ~\ref{tab:generator_ablation}, we use normal convolution, but in rows \textcircled{\scriptsize 5} to \textcircled{\scriptsize 7}, we try different combinations of DSC with our discriminators. For example, in rows \textcircled{\scriptsize 5} and \textcircled{\scriptsize 7}: DSC in MRLD and MSDFA + normal convolution in MRAD and MRPD; in row \textcircled{\scriptsize 6}: DSC in all MRLD, MSDFA, MRAD and MRPD.

We implement DSC by factorizing standard convolutions into \(K \times 1\) depthwise steps (per channel), followed by \(1 \times 1\) point-wise convolutions for cross-channel mixing. This reduces the computational complexity from \(\mathcal{O}(K\times C_{\text{in}}\times C_{\text{out}})\) to \(\mathcal{O}(K \times C_{\text{in}} + C_{\text{in}} \times C_{\text{out}})\). Here, K is  kernel size, $C_{\text{in}}$ is input and $C_{\text{out}}$ is the output channel dimension. 


\textbf{Final Design (row \textcircled{\scriptsize 7}):} Based on all these ablations, our final generator employs a total of 4 ConformerNeXt blocks (each with 8 heads and linear projection hidden dimension ×4), interconnected via Lattice Net, resolution of 5 together with DSC in MRLD and MSDFA, resolution of 3 together with normal convolution in MRAD and MRPD. This configuration achieves the best trade-off between perceptual quality (NISQA-MOS = 4.29), computational efficiency, and parameter compactness ($\approx$ 33.5 M parameters). We refer to Appendix \ref{subsec:Parameter Breakdown for Generators} and \ref{subsec:Parameter Breakdown for Discriminators} for final parameter count.

\vspace{-0.54em}
\subsection{Comparative Analysis with Baselines} 
\label{subsec:ComparativeAnalysiswithBaselines}
\vspace{-0.2em}

Table~\ref{tab:Comparative_analysis} compares our CIS-BWE against three baselines - EBEN \cite{hauret2023eben}, AERO \cite{mandel2023aero}, and AP-BWE \cite{lu2024towards} - over three extension ranges for English-VCTK and MLS French datasets. EBEN is a Pseudo Quadrature Mirror Filter-based model, AERO is a complex-valued model, and AP-BWE is a dual-stream for amplitude and phase prediction model.  




Compared to unprocessed speech in VCTK dataset, CIS-BWE significantly does a 3.3x reduction in LSD, a 1.72x increase in STOI, a 2.17x increase in PESQ, and a 1.51x increase in NISQA-MOS for 4→16 kHz. Table~\ref{tab:Comparative_analysis} also indicates that AP-BWE is the best-performing model in baselines for NISQA-MOS. Our proposed CIS-BWE exceeds AP-BWE in NISQA-MOS, LSD, PESQ, and STOI for all three frequency ranges. However, CIS-BWE gives a similar performance for SI-SDR and SI-SNR compared to AP-BWE. Please note that LSD is a measure for over-smoothing, and NISQA-MOS, PESQ, and STOI are measures of perceptual audio quality. Our model outperforms the best-performing baseline, AP-BWE, for the perceptual and over-smoothing metrics with almost 2.18x fewer parameters (72M vs 33M). This also holds for the MLS French dataset, indicating CIS-BWE outperforms AP-BWE and all other baselines in multilingual settings as well. 

\mycolor{We also calculate WER for ASR evaluation using open-source deepspeech-0.9.3-model (see Appendix \ref{sec:WERCalculation} for details). Table~\ref{tab:Comparative_analysis} indicates that  CIS-BWE can improve WER by \textit{7x} compared to unprocessed speech (i.e., 13.59\% vs.90.01\%) for 4-16 kHz extension. Moreover, CIS-BWE also outperforms AP-BWE with almost 2.18x fewer parameters (72M vs 33M) for all frequency ranges.
}

\begin{figure}[ht!]
  \centering
 \includegraphics[width=0.48\textwidth,height=0.095\textheight]{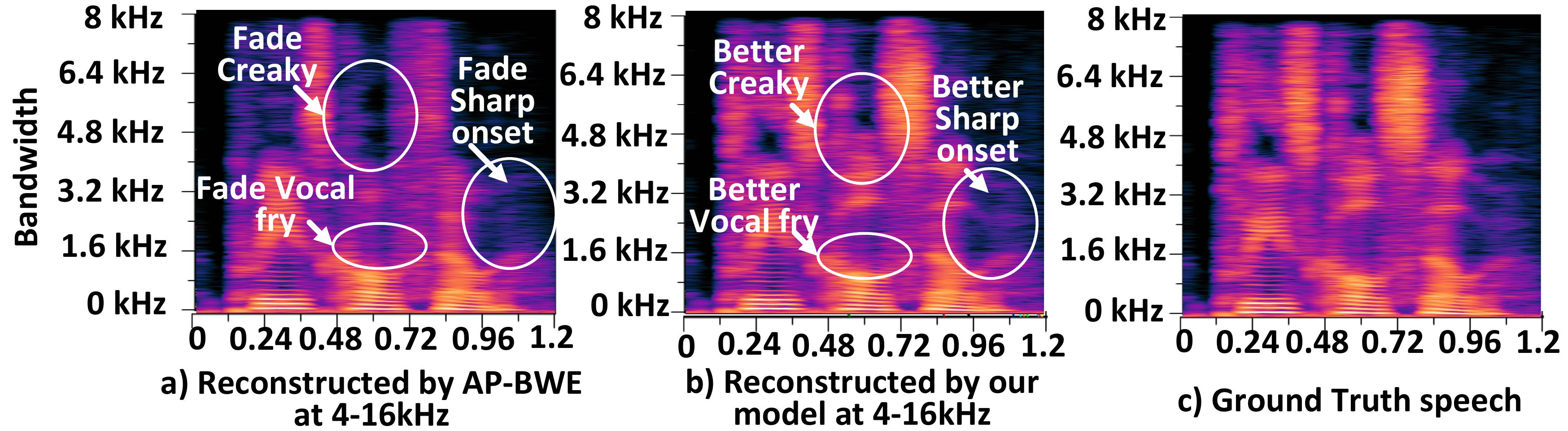} 
 \vspace{-01.805em}
  \caption{4-16 kHz extended speech by CIS-BWE. }
  \label{fig:figcomparison}
  \vspace{-0.9515em}
\end{figure}

To qualitatively illustrate the improved reconstruction performance by our CIS-BWE, we include Fig. \ref{fig:figcomparison}, which shows that chaotic cues, such as vocal fry, sharp onset, and creaky voice, are reconstructed more precisely by our CIS-BWE (almost similar to the ground truth) compared to the best-performing baseline AP-BWE. It is  an indication that our approach of chaotic modeling using chaos-informed discriminators is outperforming other nonlinear discriminators, such as MPD, with fewer parameters (see Section \ref{subsec:Comparison with MPD}).

We also refer to Table~\ref{tab:SoTA_different_fr_range} for more details on performance across different frequency ranges.

\vspace{-0.2em}

\begin{table*}[ht]
\centering
\scriptsize
\setlength{\tabcolsep}{0.85pt}
\renewcommand{\arraystretch}{0.99}

\begin{tabular}{l l l
  ccc  ccc  ccc  ccc  ccc  ccc ccc}
\toprule
\multirow{2}{*}{Method}
& \multirow{2}{*}{Size}
& \multirow{2}{*}{Data}
  & \multicolumn{3}{c}{NISQA-MOS}
  & \multicolumn{3}{c}{STOI}
  & \multicolumn{3}{c}{PESQ}
  & \multicolumn{3}{c}{SI-SDR}
  & \multicolumn{3}{c}{SI-SNR}
  & \multicolumn{3}{c}{LSD} 
  & \multicolumn{3}{c}{\mycolor{WER \%}}\\
  \cmidrule(lr){4-6}\cmidrule(lr){7-9}\cmidrule(lr){10-12}
\cmidrule(lr){13-15}\cmidrule(lr){16-18}\cmidrule(lr){19-21}\cmidrule(lr){22-24}
  & &
  & 4–16 & 8–16 & 16–48
  & 4–16 & 8–16 & 16–48
  & 4–16 & 8–16 & 16–48
  & 4–16 & 8–16 & 16–48
  & 4–16 & 8–16 & 16–48
  & 4–16 & 8–16 & 16–48 
  & \mycolor{4–16} & \mycolor{8–16} & \mycolor{16–48}\\
\midrule
\multirow{2}{*} {Unprocessed}   & \multirow{2}{*} {-}     & VCTK & 2.79 & 3.67 & 4.43 & 0.55 & 0.61 & 0.61 & 1.15  & 1.51  & 1.41  & -11.03 & -8.07 & -6.07 & -10.53 & -7.62 & -5.63 & 3.27 & 2.27 & 2.85  & \mycolor{90.0} & \mycolor{6.1} & \mycolor{2.1}\\
&   & MLS  & 2.29 & 3.12 & - & 0.51 & 0.57 & - & 1.05  & 1.34  & -  & -14.15 & -18.89 & - & -13.67 & -17.71 & - & 3.48 & 2.48 & - & \mycolor{91.1} & \mycolor{6.8} & \mycolor{-}\\

\hline

\multirow{2}{*} {EBEN, 2023}   & \multirow{2}{*} {29.7M} & VCTK & 2.59 & 2.69 & 2.53     & 0.89 & 0.98 & 0.98      & 2.64  & 3.69  & 3.71      & 11.94  & 19.94 & 20.82      & 11.94  & 19.94 & 20.83      & 1.03  & 0.78  & 0.92   & \mycolor{19.1} & \mycolor{12.4} & \mycolor{8.5} \\
   &   & MLS & 2.14 & 2.28 & -     & 0.87 & 0.96 & -      & 2.47  & 3.55  & -      & 11.81  & 18.24 & -      & 11.74  & 18.38 & -      &  1.17 &  0.88 & -   & \mycolor{19.4} & \mycolor{12.7} & \mycolor{-} \\

\hline

\multirow{2}{*} {AERO, 2023}  & \multirow{2}{*} {36.4M} & VCTK & 2.79 & 2.75 & 2.88 & 0.83 & 0.94 & 0.99  & 2.62  & 3.65  & 3.69  & 13.60  & 20.70 & 21.56 & 13.60  & 20.70 & 21.56 & 1.09  & 0.97  & 0.75 & \mycolor{20.3}  & \mycolor{13.4} & \mycolor{9.8}\\
&   & MLS  & 2.38 & 2.37 & - & 0.81 & 0.92 & -  & 2.49  & 3.45  & -  &  13.47 & 19.41 & - & 13.37  & 19.19 & - & 1.21  & 1.04  & - & \mycolor{20.7}  & \mycolor{13.7} & \mycolor{-}\\

\hline

\multirow{2}{*} {AP-BWE, 2024} & \multirow{2}{*} {72M}  & VCTK & 3.86 & 3.97 & 4.49 & 0.94 & 0.99 & 0.99  & 2.55  & 3.69  & 3.72  & 13.42  & 18.26 & 20.86 & 13.35  & 18.07 & 20.74 & 0.96  & 0.74  & 0.75 & \mycolor{13.7} & \mycolor{4.2}  & \mycolor{1.7}\\ 
&   & MLS & 3.18& 3.42 & - & 0.92&0.99  & -  & 2.51  & 3.62  & -  & 13.57  & 17.39 & - & 13.17  & 16.33 & - & 1.11  &  0.84 & - & \mycolor{13.9} & \mycolor{4.3} & \mycolor{-}\\  

\hline

\multirow{2}{*} {CIS-BWE (our)}         & \multirow{2}{*} {33.5M} & VCTK & \textbf{4.24} & \textbf{4.26} & \textbf{4.53} & \textbf{0.95} & \textbf{0.99} & \textbf{0.99}  & \textbf{2.64}  & \textbf{3.72}  & \textbf{3.75}  & \textbf{13.24}  & \textbf{18.13} & \textbf{19.53} & \textbf{13.15}  & \textbf{17.98} & \textbf{19.44} & \textbf{0.95}  & \textbf{0.72}  & \textbf{0.71} & \mycolor{\textbf{13.5}} & \mycolor{\textbf{3.9}} & \mycolor{\textbf{1.5}}\\
&   & MLS & \textbf{3.47} & \textbf{3.91} & - & \textbf{0.92} & \textbf{0.99} & -  & \textbf{2.58}  & \textbf{3.71} & -  & \textbf{13.29}  & \textbf{17.68} & - & \textbf{12.86}  & \textbf{16.81} & - & \textbf{1.10}  & \textbf{0.82}  & - & \mycolor{\textbf{13.8}} & \mycolor{\textbf{4.1}} & \mycolor{\textbf{-}}\\ 

\bottomrule
\end{tabular}

\vspace{-01.3em}
\caption{\mycolor{Comparison for clean English-VCTK and MLS French dataset (MLS does not have 48 kHz speech).}}
\label{tab:Comparative_analysis}
\vspace{-02.493em}
\end{table*}

\vspace{-0.8705em}

\vspace{-0.86em}
\subsection{Study for Noisy Conditions} 
\label{subsec:Analysis for Different Frequency Ranges}
\vspace{-0.2em}

We use eight different noise sources from the AURORA dataset \cite{pearce00_icslp} and use five different SNRs: -10, -5, 0, 5, and 10 dB for each of the noise sources to separately create the noisy dataset from the clean VCTK and MLS French. The eight noise sources are chosen from stationary and non-stationary types, such as airports, babble, cars, exhibitions, train stations, city streets, speech-shaped noise, and white Gaussian noise. 
We test our model separately for English and French noisy datasets. The average results for all noise sources and SNRs are shown in Table \ref{tab:noiseComparative_analysis}, indicating that CIS-BWE outperforms the best-performing baseline AP-BWE.  

\vspace{-0.830em}
\begin{table}[ht]
\centering
\scriptsize
\setlength{\tabcolsep}{1.2pt}
\renewcommand{\arraystretch}{0.85}
\begin{tabular}{l l 
  ccc  ccc  ccc}
\toprule
\multirow{2}{*}{Method}
& \multirow{2}{*}{Data}
  & \multicolumn{3}{c}{NISQA-MOS}
  & \multicolumn{3}{c}{SI-SNR}
  & \multicolumn{3}{c}{LSD} \\
\cmidrule(lr){3-5}\cmidrule(lr){6-8}\cmidrule(lr){9-11}
  &
  & 4–16 & 8–16 & 16–48
  & 4–16 & 8–16 & 16–48
  & 4–16 & 8–16 & 16–48 \\
\midrule

\multirow{2}{*} {EBEN}   & MLS & 0.84 &  0.92 & -  & 4.12  & 5.20 &   -   & 1.43  & 1.13 &  -   \\
 & VCTK & 1.01 & 1.08 & 1.15 & 4.23  & 5.31 & 6.01    & 1.41  & 1.11  & 1.00     \\

\hline
\multirow{2}{*} {AERO}  & MLS & 1.05 & 1.09 &  - & 4.21  & 5.24 & - & 1.45 & 1.16  & - \\
 & VCTK & 1.52 & 1.12 & 1.19 & 4.39 & 5.38  & 6.15 & 1.43 & 1.13  &  1.01  \\

\hline
\multirow{2}{*} {AP-BWE}  & MLS & 2.01 & 2.19 & -  & 4.10  & 5.14 & - &  1.42 & 1.17  & - \\
 & VCTK  & 2.74 & 3.15 & 3.71  & 4.36  & 5.31 & 6.11 & 1.37  & 1.07  &  0.89\\

\hline
\multirow{2}{*} {CIS-BWE}        & MLS & \textbf{3.14} & \textbf{3.24} & \textbf{-}  & \textbf{4.51}  & \textbf{5.71} & \textbf{-} & \textbf{1.40}  & \textbf{1.15}  & \textbf{-} \\
        & VCTK & \textbf{3.72} & \textbf{3.79} & \textbf{3.89} & \textbf{4.52}  & \textbf{5.75} & \textbf{6.22} & \textbf{1.35}  & \textbf{1.06}  & \textbf{0.88} \\
\bottomrule
\end{tabular}

\vspace{-01.3em}
\caption{Analysis for noisy VCTK and MLS dataset.}
\label{tab:noiseComparative_analysis}
\vspace{-02.493em}
\end{table}

\vspace{-0.2em}
\subsection{SNR-Wise Noise Analysis} 
\label{subsec:SNR-Wise Noise Analysis}
\vspace{-0.2em}

Table \ref{tab:noiseComparative_analysis_forSNR} indicates that our proposed CIS-BWE outperforms the best-performing baseline AP-BWE for all noisy SNRs, especially for NISQA-MOS, indicating that CIS-BWE generates better quality speech in noisy scenarios. It is intuitive that the noisiest SNR of -10 dB gives the worst result, and the least noisy SNR of 10 dB gives the best performance. We refer to Appendix \ref{sec:noisydataset_dataloader} and \ref{sec:noisenalaysis} for more details on noise type- and SNR-wise analysis.

\vspace{-0.30em}
\begin{table}[ht!]
\centering
\scriptsize
\setlength{\tabcolsep}{1.2pt}
\renewcommand{\arraystretch}{0.85}
\begin{tabular}{l l 
  ccc  ccc  ccc}
\toprule
\multirow{2}{*}{SNR}
& \multirow{2}{*}{Model}
  & \multicolumn{3}{c}{NISQA-MOS}
  & \multicolumn{3}{c}{SI-SNR}
  & \multicolumn{3}{c}{LSD} \\
\cmidrule(lr){3-5}\cmidrule(lr){6-8}\cmidrule(lr){9-11}
  &
  & 4–16 & 8–16 & 16–48
  & 4–16 & 8–16 & 16–48
  & 4–16 & 8–16 & 16–48 \\
\midrule
\multirow{2}{*} {-10}  & AP-BWE     & 2.02 & 2.55 &  2.98 & -0.60  & -0.04 & 1.17 & 1.49 & 1.18 & 0.91 \\
  & CIS-BWE     & \textbf{2.73} & \textbf{3.05} & \textbf{3.37}  & \textbf{0.22} & \textbf{1.09} & \textbf{1.21} & \textbf{1.45} & \textbf{1.18} & \textbf{0.90} \\

\hline
\multirow{2}{*} {-5}   & AP-BWE & 2.52 &  2.86 & 3.54 &  2.86 & 3.48 &  4.44   & 1.45 & 1.11 &  0.88    \\
 & CIS-BWE & \textbf{3.56} & \textbf{3.51} & \textbf{3.73} & \textbf{3.35}  & \textbf{4.23} & \textbf{4.01}     & \textbf{1.54}  & \textbf{1.12} & \textbf{0.90}     \\

\hline
\multirow{2}{*} {0}  & AP-BWE & 2.83 & 3.21 & 3.80  &  5.06 & 5.93 & 6.80 & 1.39  &  1.06 & 0.86 \\
  & CIS-BWE  & \textbf{3.90} & \textbf{3.91} & \textbf{3.96}  & \textbf{5.17}  & \textbf{6.31} & \textbf{6.80} & \textbf{1.34}  & \textbf{1.05}  & \textbf{0.85} \\

\hline
\multirow{2}{*} {5}  & AP-BWE & 3.05 & 3.42 & 4.07  & 6.53  & 7.74 & 8.64 & 1.34  & 1.02  & 0.84 \\
 & CIS-BWE  & \textbf{4.15} & \textbf{4.08} & \textbf{4.12}  & \textbf{6.38}  & \textbf{7.87} & \textbf{9.18} & \textbf{1.32}  & \textbf{1.01}  &  \textbf{0.82}\\

\hline
\multirow{2}{*} {10}        & AP-BWE & 3.22 & 3.70 & 4.22  & 7.62  & 9.02 & 9.77 & 1.29  & 0.97  & 0.82 \\
        & CIS-BWE & \textbf{4.27} & \textbf{4.29} & \textbf{4.31} & \textbf{7.17}  & \textbf{8.92} & \textbf{10.90} & \textbf{1.28}  & \textbf{0.96}  & \textbf{0.81} \\
\bottomrule
\end{tabular}

\vspace{-01.0em}
\caption{SNR-wise performance for noisy VCTK.}
\label{tab:noiseComparative_analysis_forSNR}
\end{table}

\vspace{-01.96em}
\subsection{Computational Complexity}
\label{subsec:Computational Complexity}
\vspace{-0.32em}

Table \ref{tab:computation} shows the computational complexity  using Multiply Accumulate Operations (MACs), Floating Point Operations per second (FLOPs), Real-Time Factor (RTF), and inference time. \mycolor{Due to optimization of generators and discriminators (see Sections \ref{subsec:generator_ablation}, \ref{subsec:Discriminator_Efficient}, \ref{subsec:Parameter Breakdown for Discriminators}, \ref{subsec:Parameter Breakdown for Generators}), CIS-BWE uses 0.5x fewer parameters, MACs, and FLOPs compared to AP-BWE, meaning  $\sim$0.5x less memory usage than AP-BWE while maintaining superior perceptual quality as shown by NISQA-MOS in Table~\ref{tab:Comparative_analysis}.}  CIS-BWE also has a lower/similar RTF and inference time, indicating its effectiveness in real-time services (see Appendix \ref{append:analysisoncomputationcomplexity} for detailed analysis). 

\vspace{-0.6em}
\begin{table}[ht]
\centering
\scriptsize
\setlength{\tabcolsep}{1.6pt}
\renewcommand{\arraystretch}{0.99}
\begin{tabular}{l|c|c|c|c|c|c}
\toprule
Model & Freq. & Par.(M) & MAC(M) & FLOP(M) & RTF(GPU) & Inf.(ms) \\
\hline
AP-BWE & 4-16 kHz & 72.07& 14236.65 & 28473.31  & 0.0023x & 14.93 \\
AP-BWE & 16-48 kHz & 72.07& 14236.65 & 28473.31 &  0.0025x & 16.60 \\
CIS-BWE & 4-16 kHz & \textbf{33.74} & \textbf{6790.86} & \textbf{13581.73} & \textbf{0.0025x} & \textbf{13.60}\\
CIS-BWE & 16-48 kHz & \textbf{33.74} & \textbf{6790.86} & \textbf{13581.73} & \textbf{0.0028x} & \textbf{16.64}\\
\bottomrule
\end{tabular}
\vspace{-0.94em}
\caption{Computational complexity of CIS-BWE. The hardware configuration is provided in Section \ref{subsec:Hyperparameter_Configuration}.}
\label{tab:computation}
\vspace{-01.4em}
\end{table}


\vspace{-0.87em}
\subsection{Subjective Analysis}
\label{subsec:Subjective Test}
\vspace{-0.2em}

\begin{figure}[ht!]
  \centering
 \includegraphics[width=0.48\textwidth,height=0.11\textheight]{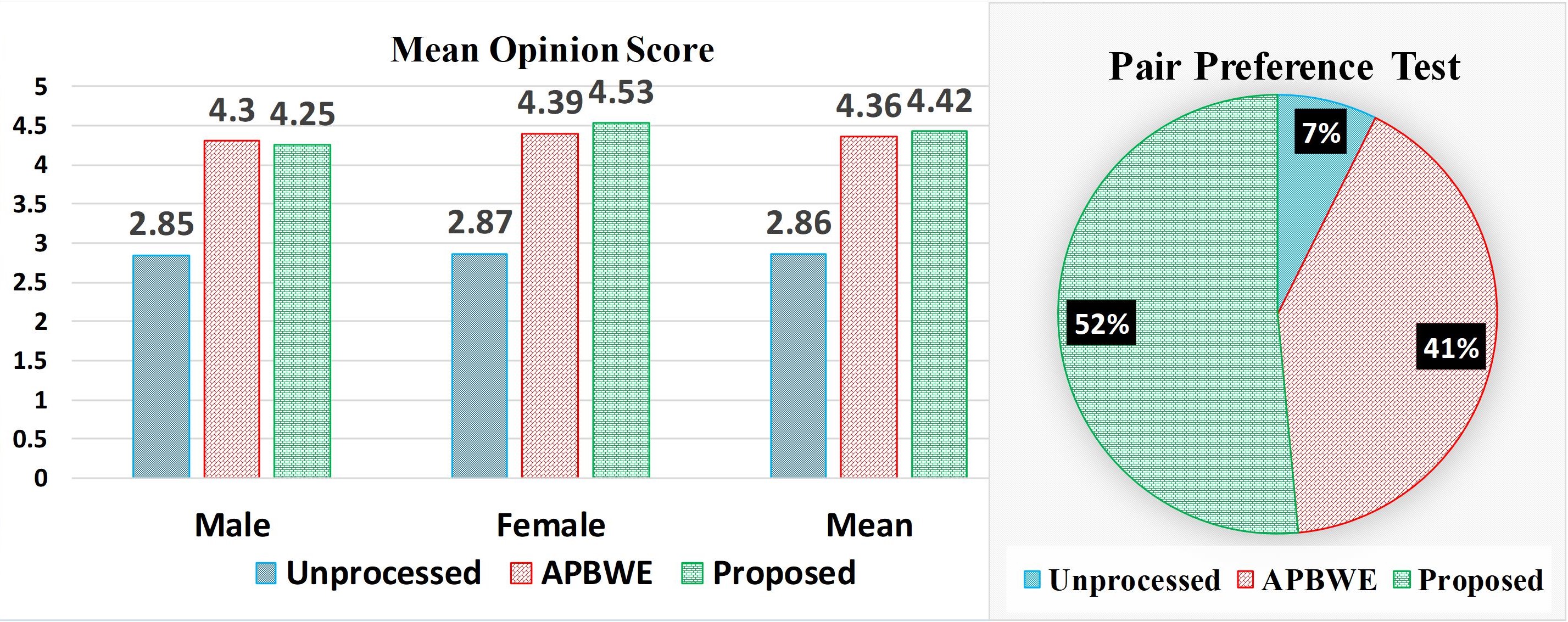} 
 \vspace{-01.905em}
  \caption{Results of MOS and Pairwise preference test. }
  \label{fig:comparison}
  \vspace{-0.9515em}
\end{figure}

\mycolor{We use 5-point (1=bad to 5=excellent) Mean Opinion Score (MOS) ratings and Pairwise preference tests for subjective comparison using 10 volunteers (see Appendix \ref{subsec:subj_eval_details} for details).} In Fig. \ref{fig:comparison}, we present the MOS results separately for male and female speakers with the overall mean. AP-BWE performs better for only male speakers, while CIS-BWE outperforms for female speakers and overall. In the Pairwise preference test, CIS-BWE outperforms AP-BWE by a margin of 11\%. 
These results provide strong evidence that our  CIS-BWE consistently generates higher perceptual quality audio, which is favored by a wide range of listeners.     

\vspace{-0.9705em}

\vspace{-0.974em}
\section{Conclusion} 
\vspace{-0.3em}

We propose CIS-BWE, an adversarial model for speech BWE. To the best of our knowledge, for the first time, we incorporate chaotic dynamics of speech for improved perceptual quality in a dual-stream GAN-based framework. The efficacy of CIS-BWE is shown across a wide range of performance metrics. We believe that our chaos-informed discriminators will be adopted in the future in a wide range of NLP applications in the domain of generative speech for TTS and ASR tasks.

\vspace{-0.8em}
\section{Acknowledgment} 

We are thankful to the VCTK and MLS data collection team for creating and releasing the VCTK and MLS corpus and the AP-BWE authors \cite{lu2024towards} for providing their GitHub codebase, which we have substantially modified and extended for this work. The codebase is under MIT license and open source. We also express gratitude to the participants of the subjective evaluation tests for their contributions to the listening tests. We acknowledge that we have used Elicit for finding relevant papers, and used ChatGPT for debugging codes, and finding grammatical errors.



\section{Limitations}


\mycolor{The training time/epoch for CIS-BWE is slightly higher than the SOTA AP-BWE (17 minutes/epoch vs 25 minutes/epoch) because of the involved large number of chaotic features, controlled cross-stream mixing, and nonlinear calculations.}

\mycolor{For better understanding of the slightly larger epoch time, we calculate the computational overhead for the Lyapunov Exponents (LE) and Detrended Fractal (DF) features, which are applied to the raw waveform before the discriminator's CNN backbone. We calculate MACs, FLOPs, and latency for each segment (i.e., 8000 samples) that are shown in Table \ref{tab:epochtimeComparative_analysis}  for LE and DF features.}

\vspace{-0.630em}
\begin{table}[ht]
\centering
\scriptsize
\setlength{\tabcolsep}{01.2pt}
\renewcommand{\arraystretch}{0.85}
{\color{black}
\begin{tabular}{ 
  ccc  ccc}
\toprule
   \multicolumn{3}{c}{Lyapunov Exponents (LE)}
  & \multicolumn{3}{c}{Detrended Fractal (DF)}\\
\cmidrule(lr){1-3}\cmidrule(lr){4-6}

   MACs (M) & FLOPs (M) & Latency (ms)
   & MACs (M) & FLOPs (M) & Latency (ms)\\
\midrule
 707.7 &  1415.5 & 6.56  & 0.277  & 0.554 &   0.145     \\
\bottomrule
\end{tabular}
}
\vspace{-0.83em}
\caption{\mycolor{Computational complexity analysis only for LE and DF calculations in MRLD and MSDFA.}}
\label{tab:epochtimeComparative_analysis}
\vspace{-01.493em}
\end{table}

\mycolor{Table \ref{tab:epochtimeComparative_analysis} indicates that the 8-minute training-time overhead (25 min vs 17 min) is primarily due to LE estimation in MRLD, as DF overhead in MSDFA is negligible. However, the MACs and FLOPs overhead for LE is only 10.41\% of the total MAC of 6790.86 M (see Table \ref{tab:computation}). \textit{Please also note that as LE and DF are only present in discriminators, their presence does not impact the inference time.} Moreover, our inference time is 1.2x smaller than the best-performing model AP-BWE (see Table \ref{tab:computation}) with $\approx 0.5\times$ fewer parameters, MACs, and FLOPs than AP-BWE which makes CIS-BWE a new baseline for the BWE task. We refer to Appendix \ref{append:analysisoncomputationcomplexityepochtime} for more detailed analysis on this.}




\section{Potential Risks / Ethical Considerations} 

While the intention of designing the CIS-BWE model is for frequency restoration research purposes, it can be misused for potential secret eavesdropping, impersonation, or deepfake audio generation. This model has the potential for severe privacy and security risks. To avoid these, we have to be very careful to ensure transparency, protect consent, and always follow guidelines.

\bibliography{custom}

\appendix

\section{Appendix}
\label{sec:appendix}

\subsection{Chaotic Properties of Speech Generation}
\label{subsec:Chaotic_PropertiesofSpeechGeneration}

Speech production is fundamentally a \emph{non-linear dynamical process characterized by deterministic chaos} \cite{jiang2001modeling}, \cite{fitch2025applying}. Its generation is driven by aerodynamic forces, with visco-elastic vocal cords forming a self-sustained oscillatory system whose glottal pulses produce harmonic frequencies, pressure waves occurring at integer multiples of the fundamental frequency (f0) \cite{titze2008nonlinear}. According to the source-filter theory, these harmonics are then shaped by the vocal tract's filtering action—resonances in the throat, mouth, and nasal cavities dynamically amplify or attenuate certain harmonics, forming moving spectral peaks known as \emph{formants} \cite{zhang2023influence}.

Because the vocal cords receive pressure feedback from the vocal tract, this coupled system can undergo period-doubling, create sub-harmonic frequencies, and intermittently exhibit \emph{chaotic behavior} indicated by positive Lyapunov exponents \cite{martinez2002detection}. Consequently, speech naturally alternates between stable, quasi-periodic sounds (typical vowels) and chaotic segments, such as creaky voice and stressed speech, resulting in subtle jittery fluctuations, turbulence, and irregular timing that purely linear models cannot capture. Generally, vowels exhibit quasi-periodicity interspersed with intermittent chaotic episodes \cite{tao2008chaotic}.

During sound excitation, unstable airflow, vortex shedding, and uneven vocal cord movements introduce additional turbulence and timing irregularities. Within the vocal tract, constricted passages like those forming fricatives produce local turbulence, while the reactive characteristics of supraglottal and subglottal airways feed pressure variations back to the vocal cords, creating a complex non-linear interaction that can either stabilize or destabilize cord oscillations \cite{may2023effects}. Moderate coupling enriches harmonic content and clarifies formant structures, whereas strong coupling can induce chaotic behaviors, resulting in rough or harsh vocal qualities as observed in creaky voices, infant cries, or animal distress calls \cite{rendall2025nonlinear}.

A purely linear model overlooks critical aspects such as sub-harmonics, bifurcations, and aperiodic bursts, making synthetic speech sound unnaturally smooth \cite{sheng2019reducing}. Moreover, diagnostic methods that rely on detecting chaotic indicators for early identification of vocal disorders would lose effectiveness. Contemporary speech synthesis and enhancement systems predominantly use linear models or perturbation parameters, failing to capture these complex, subtle dynamics of speech \cite{maccallum2009acoustic}. Hence, to accurately represent these non-linear behaviors, models must incorporate non-linear glottal-flow representations or leverage adversarial networks with discriminators designed to recognize non-linear characteristics, ensuring both the deterministic aspects (such as harmonic and formant structures) and chaotic elements (including noise bursts and timing irregularities) are faithfully reproduced \cite{bollepalli2019generative}.

\subsection{WER Calculation for ASR Tasks}
\label{sec:WERCalculation}

\mycolor{We explicitly evaluate CIS-BWE for downstream ASR tasks. Following prior work on silent-speech voicing \cite{gaddy2021improved} and the Mozilla DeepSpeech architecture \cite{hannun2014deep}, we use the open-source \texttt{deepspeech-0.9.3-models} checkpoint to convert both band-limited and CIS-BWE--enhanced audios into text, and we compute word error rate (WER), character error rate (CER), and word accuracy (WAcc) using the same procedure as \cite{gaddy2021improved}. CIS-BWE consistently reconstructs missing high frequency contents which transform degraded band-limited speech into signals suitable for ASR. For example, in Table~\ref{tab:ASR_eval}, for the 4--16~kHz condition, the CER drops from $68.59\%$ (Unprocessed) to $7.72\%$, WER drops from $90.01\%$ (Unprocessed) to $13.59\%$, and WAcc improves from $10.12\%$ (Unprocessed) to $86.50\%$. We also see significant improvement in all the frequency ranges. These improvements clearly demonstrate that CIS-BWE preserves phonetic cues that are directly useful for downstream recognition. }

\vspace{-0.650em}
\begin{table}[ht!]
\centering
\scriptsize
\setlength{\tabcolsep}{3pt}
\renewcommand{\arraystretch}{0.85}
{\color{black}
\begin{tabular}{lcccccc}
\toprule
Model &  Range & WER  & CER  & Word Accuracy\\
\midrule
Unprocessed & 4 kHz & 90.01\% & 68.59\% & 10.12\% \\
CIS-BWE & 4-16 kHz & 13.59 \% & 7.72\% & 86.50\% \\
Unprocessed & 8 kHz & 6.14\% & 3.15\% & 93.96\%\\
CIS-BWE & 8-16 kHz & 3.9\% & 1.5\% & 96.14\% \\
Unprocessed & 16 kHz & 2.1\% & 1.21\% & 97.47\%\\
CIS-BWE & 16-48 kHz & 1.5\% & 0.79\% & 98.49\%\\

\bottomrule
\end{tabular}
}
\vspace{-0.94em}
\caption{\mycolor{WER on the VCTK dataset.}}
\label{tab:ASR_eval}
\vspace{-02.84em}
\end{table}

\subsection{Dataset Pre‐Processing Pipeline}
\label{sec:dataset_preprocessing} 

We use VCTK dataset, which is an established benchmark extensively use in Speech processing tasks. We extensively check and ensure that VCTK dataset does not contain any Personal Identifying Information (PIN), abusive contents, or any harmful that might be harmful for any individual, group, or others. We at first index all audio files by reading each line from \texttt{training.txt} and \texttt{test.txt} files, by extracting base filenames (without extensions) and by splitting on the “\texttt{|}” character. Out of 110 English native speakers with different accents reading Herald Newspaper articles, we have used 102 speakers for training the CIS-BWe and 8 speakers for testing the efficacy of the CIS-BWE. In total we have used 88,329 audio recordings for training and testing the CIS-BWE. For reducing disk I/O overhead, each audio cache in memory for up to \texttt{n\_cache\_reuse} accesses (default set to 1). Then we trim the silent portion of the audio files by deleting portion of audios taking the start and end of silences from the \texttt{vctk-silences.0.92.txt}. After that, audios are loaded by \texttt{torchaudio.load}, and stereo (dual) channel signals are converted into mono (single) by averaging across channels. These mono waveforms are then resampled to the high‐resolution (HR) target of 16/48 KHz by sinc interpolation if requires. Furthermore, a low‐resolution (LR) version is created by first downsampling the audio to 2/4/8/16/24 KHz and then upsampled back to 16/48 KHz by sinc interpolation which is given as input to the CIS-BWE model. For using the audios in training mode (\texttt{split=True}), we randomly crop an 8{,}000‐sample segment, which is approximately 167 ms from HR and LR signals both. If the files are shorter than this length, then zero‐padding is applied to ensure similar segment lengths.

For MLS French dataset, we use similar pre-processing as we do for VCTK. We use the default validation and test split as given in the corpus. For training set, at first we use the default 9 hours provided by the dataset. However, due to smaller subset of the dataset, CIS-BWE performs poorly due to sub-optimal training. 
Therefore, we choose the full dataset of $\sim$1000 hours to train the models.

\subsection{Noisy Dataset Preparation}
\label{sec:noisydataset_dataloader}

We have collected noises from AURORA dataset. We choose eight different types of noise and 5 different SNR levels to mix with the clean speech to simulate real world environment. We continuously concatenate the same noise file back-to-back in a loop until each file is 5 minutes long without any cross-fades. For each clean speech file, we randomly choose one noise type and one SNR out of 40 different possible combinations. Then we randomly crop a segment exactly the same length as the clean audio. Then we scale the noise to fulfill the chosen SNR and add it with the clean audio. A CSV log file is used to track the noise type, the SNR, the start-end index for the crop for reproducibility.

\subsection{SNR- and Noise type-Wise Noise Analysis in Details}
\label{sec:noisenalaysis}

\textbf{Table \ref{tab:Comparative_analysis_SNR_wise_CISBWE} (CIS-BWE):} Performance increases significantly as SNR increases from $-10$ to $10$ dB, with consistent gains in NISQA-MOS, STOI, PESQ, SI-SDR, and SI-SNR, and a corresponding drop in LSD. The widest band (16–48 kHz) yields the \emph{strongest} SI-SDR/SI-SNR improvements and the \emph{lowest} LSD at high SNRs—evidence which shows that richer high-frequency cues are being faithfully reconstructed  in the presence of noises due to the non-linear feedback from chaos discriminators. Overall, CIS-BWE remains robust even at $-10$ dB and performance scales linearly to $10$ dB.

\textbf{Table \ref{tab:Comparative_analysis_snr_wise_apbwe} (AP-BWE):} AP-BWE follows the same linear SNR trend, but its absolute quality and intelligibility lags CIS-BWE across bands, with the gap most visible under low SNRs and wider extensions.

CIS-BWE also consistently performs robustly in Tables \ref{tab:Comparative_analysis_noise_SNR_wise_cisbwe} and \ref{tab:Comparative_analysis_noise_source_wise-APBWE}, where results are showed according to the noise types. CIS-BWE performs better in real-world noises such as Exhibition, Airport, and Street noises. AP-BWE also follows a similar pattern; however, CIS-BWE outperforms AP-BWE.

\begin{table*}[!ht]
\centering
\scriptsize
\setlength{\tabcolsep}{1.8pt}
\renewcommand{\arraystretch}{0.85}

\begin{tabular}{l ccc ccc ccc ccc ccc ccc ccc}
\toprule
\multirow{2}{*}{SNR}
& \multicolumn{3}{c}{Samples}
& \multicolumn{3}{c}{NISQA-MOS}
& \multicolumn{3}{c}{STOI}
& \multicolumn{3}{c}{PESQ}
& \multicolumn{3}{c}{SI-SDR}
& \multicolumn{3}{c}{SI-SNR}
& \multicolumn{3}{c}{LSD} \\
\cmidrule(lr){2-4}\cmidrule(lr){5-7}\cmidrule(lr){8-10}\cmidrule(lr){11-13}\cmidrule(lr){14-16}\cmidrule(lr){17-19}\cmidrule(lr){20-22}
& 4--16 & 8--16 & 16--48
& 4--16 & 8--16 & 16--48
& 4--16 & 8--16 & 16--48
& 4--16 & 8--16 & 16--48
& 4--16 & 8--16 & 16--48
& 4--16 & 8--16 & 16--48
& 4--16 & 8--16 & 16--48 \\
\midrule
 
-10  & 558 & 558 & 593 & 2.73 & 3.05 & 3.375 & 0.64 & 0.74 & 0.73 & 1.11  & 1.23  & 1.16 & 0.29 & 1.17 &  0.67 & 0.22  &  1.09 & 0.61  & 1.558 & 1.18 &  0.93 \\
-5   & 585 & 585 & 608 & 3.56 & 3.51 &  3.73 & 0.73 & 0.82 & 0.8  & 1.19  &  1.45 &  1.25& 3.47 & 4.35 &  4.13 &  3.35 &  4.23 &  4.01 & 1.54  & 1.12 &   0.9   \\
0    & 606 & 606 & 601 & 3.9  & 3.91 & 3.96  & 0.79 & 0.87 & 0.85 & 1.28  &  1.71 &  1.4 & 5.12 & 6.38 &  6.9  & 5.17  &  6.31 &  6.8  & 1.49  & 1.07 &  0.88 \\
5    & 580 & 580 & 576& 4.15 &  4.08 & 4.12  & 0.83 & 0.91 & 0.89 & 1.38  &  2.04 & 1.6  & 6.44 & 7.93 &  9.28 & 6.38  &  7.87 &  9.18 & 1.43  & 1.03 &  0.86  \\
10   & 608 & 608 & 559& 4.27 &  4.29 & 4.31 & 0.86 & 0.94 & 0.93  & 1.50 &  2.35 &  1.84 & 7.18 & 8.93 &  10.95& 7.17  & 8.92  &  10.9& 1.38 &  0.99 &  0.84  \\
\bottomrule
\end{tabular}

\vspace{-01.3em}
\caption{Comparative SNR-wise analysis over three extension ranges of our proposed CIS-BWE on VCTK noisy dataset.}
\label{tab:Comparative_analysis_SNR_wise_CISBWE}
\vspace{-01.53em}
\end{table*}

\begin{table*}[!ht]
\centering
\scriptsize
\setlength{\tabcolsep}{1.8pt}
\renewcommand{\arraystretch}{0.85}

\begin{tabular}{l ccc ccc ccc ccc ccc ccc ccc}
\toprule
\multirow{2}{*}{Noise Types}
& \multicolumn{3}{c}{Samples}
& \multicolumn{3}{c}{NISQA-MOS}
& \multicolumn{3}{c}{STOI}
& \multicolumn{3}{c}{PESQ}
& \multicolumn{3}{c}{SI-SDR}
& \multicolumn{3}{c}{SI-SNR}
& \multicolumn{3}{c}{LSD} \\
\cmidrule(lr){2-4}\cmidrule(lr){5-7}\cmidrule(lr){8-10}\cmidrule(lr){11-13}\cmidrule(lr){14-16}\cmidrule(lr){17-19}\cmidrule(lr){20-22}
& 4--16 & 8--16 & 16--48
& 4--16 & 8--16 & 16--48
& 4--16 & 8--16 & 16--48
& 4--16 & 8--16 & 16--48
& 4--16 & 8--16 & 16--48
& 4--16 & 8--16 & 16--48
& 4--16 & 8--16 & 16--48 \\
\midrule
AWGN       & 348 & 348 & 378 & 3.73 & 3.34    &  3.23   & 0.77 &  0.82   &  0.82   & 1.29 & 1.49    &  1.40   & 5.13 &  6.17   &  7.25   & 5.04 & 6.09    &  7.19   & 1.49 &  1.35   &  0.90 \\
Airport    & 372 & 372 & 411 & 3.77 & 3.92    &  4.16   & 0.77 &  0.87   &  0.85   & 1.31 & 1.84    &  1.49   & 4.60 &  5.87   &  6.28   & 4.56 & 5.84    &  6.19   & 1.48 &   1.01  &  0.85   \\
Babble     & 384 & 384 & 348 & 3.67 & 3.9    &  4.07   & 0.76 &   0.87  &  0.85   & 1.28 &  1.79   &  1.46   & 4.18 &   5.62  &  6.02   & 4.15 & 5.59    &  5.93   & 1.47 &   1.01  &  0.86  \\
Car        & 354 & 354 & 370 & 3.75 & 3.86    &  4.04   & 0.78 &  0.87   &  0.85   & 1.29 & 1.81    &  1.46   & 4.87 &  5.99   &  6.33   & 4.79 & 5.93    &  6.27   & 1.50 &   1.03  &  0.88   \\
Exhibition & 333 & 333 & 366 & 3.94 & 3.97    &  4.26   & 0.80 &  0.88   &  0.85   & 1.33 & 1.93    &  1.47   & 5.39 &  6.61   &  7.19   & 5.34 & 6.57    &  7.09   & 1.46 &   1.01  &  0.87   \\
SSN        & 386 & 386 & 359 & 3.49 & 3.59    &  3.26   & 0.77 &  0.84   &  0.82   & 1.32 & 1.68    &  1.37   & 3.46 &  4.66   &  4.78   & 3.37 & 4.55    &  4.63   & 1.46 &   1.15  &  0.99   \\
Station    & 380 & 380 & 349 & 3.59 & 3.78    &  4.12   & 0.76 &  0.86   &  0.85   & 1.27 & 1.78    &  1.48   & 4.31 &  5.64   &  6.29   & 4.20 & 5.54    &  6.18   & 1.49 &   1.01  &  0.85   \\
Street     & 380 & 380 & 356 & 3.81 & 3.88    &  4.01   & 0.78 &  0.86   &  0.84   & 1.30 & 1.81    &  1.42   & 4.93 &  6.15   &  6.27   & 4.88 & 6.09    &  6.20   & 1.48 &   1.03  &  0.86   \\
\bottomrule
\end{tabular}

\vspace{-1.3em}
\caption{Comparative noise type–wise analysis over three extension ranges of our proposed CIS-BWE on the VCTK noisy dataset.}
\label{tab:Comparative_analysis_noise_SNR_wise_cisbwe}
\vspace{-1.493em}
\end{table*}

\begin{table*}[!ht]
\centering
\scriptsize
\setlength{\tabcolsep}{1.8pt}
\renewcommand{\arraystretch}{0.85}

\begin{tabular}{l ccc ccc ccc ccc ccc ccc ccc}
\toprule
\multirow{2}{*}{SNR}
& \multicolumn{3}{c}{Samples}
& \multicolumn{3}{c}{NISQA-MOS}
& \multicolumn{3}{c}{STOI}
& \multicolumn{3}{c}{PESQ}
& \multicolumn{3}{c}{SI-SDR}
& \multicolumn{3}{c}{SI-SNR}
& \multicolumn{3}{c}{LSD} \\
\cmidrule(lr){2-4}\cmidrule(lr){5-7}\cmidrule(lr){8-10}\cmidrule(lr){11-13}\cmidrule(lr){14-16}\cmidrule(lr){17-19}\cmidrule(lr){20-22}
& 4--16 & 8--16 & 16--48
& 4--16 & 8--16 & 16--48
& 4--16 & 8--16 & 16--48
& 4--16 & 8--16 & 16--48
& 4--16 & 8--16 & 16--48
& 4--16 & 8--16 & 16--48
& 4--16 & 8--16 & 16--48 \\
\midrule
 
-10  & 558 & 558 & 593 & 2.02 & 2.55  & 2.98 & 0.60  &  0.72 & 0.73 & 1.08  & 1.19  & 1.15 & -0.52 & 0.05 & 1.24 & -0.60 & -0.04  & 1.17  & 1.49 & 1.18 &  0.91 \\
-5   & 585 & 585 & 608 & 2.52 & 2.86  & 3.54 & 0.69  & 0.80  & 0.80 & 1.15  & 1.38  & 1.26 &  2.96 & 3.62 & 4.54 &  2.86 & 3.48   & 4.44  & 1.45 & 1.11 &  0.88    \\
0    & 606 & 606 & 601 & 2.83 & 3.21  & 3.80 & 0.76  &  0.86 & 0.85 & 1.24  & 1.63  & 1.41 &  5.13 & 6.03 & 6.87 &  5.06 & 5.93   & 6.80  & 1.39 & 1.06 &  0.86\\
5    & 580 & 580 & 576 & 3.05 &  3.42 & 4.07 & 0.81  &  0.90 & 0.89 & 1.35  & 1.93  & 1.61 &  6.57 & 7.81 & 8.68 &  6.53 & 7.74   & 8.64  & 1.34 & 1.02 &  0.84  \\
10   & 608 & 608 & 559 & 3.22 &  3.70 & 4.22 & 0.84  &  0.93 & 0.92 & 1.49  & 2.26  & 1.80 &  7.63 & 9.05 & 9.77 &  7.62 & 9.02   & 9.77  & 1.29 & 0.97 &  0.82 \\
\bottomrule
\end{tabular}

\vspace{-01.3em}
\caption{Comparative SNR-wise analysis over three extension ranges of our proposed AP-BWE on VCTK noisy dataset.}
\label{tab:Comparative_analysis_snr_wise_apbwe}
\vspace{-01.53em}
\end{table*}

\begin{table*}[!ht]
\centering
\scriptsize
\setlength{\tabcolsep}{1.8pt}
\renewcommand{\arraystretch}{0.85}

\begin{tabular}{l ccc ccc ccc ccc ccc ccc ccc}
\toprule
\multirow{2}{*}{Noise Types}
& \multicolumn{3}{c}{Samples}
& \multicolumn{3}{c}{NISQA-MOS}
& \multicolumn{3}{c}{STOI}
& \multicolumn{3}{c}{PESQ}
& \multicolumn{3}{c}{SI-SDR}
& \multicolumn{3}{c}{SI-SNR}
& \multicolumn{3}{c}{LSD} \\
\cmidrule(lr){2-4}\cmidrule(lr){5-7}\cmidrule(lr){8-10}\cmidrule(lr){11-13}\cmidrule(lr){14-16}\cmidrule(lr){17-19}\cmidrule(lr){20-22}
& 4--16 & 8--16 & 16--48
& 4--16 & 8--16 & 16--48
& 4--16 & 8--16 & 16--48
& 4--16 & 8--16 & 16--48
& 4--16 & 8--16 & 16--48
& 4--16 & 8--16 & 16--48
& 4--16 & 8--16 & 16--48 \\
\midrule
AWGN       & 348 & 348 & 378 & 2.68  &  2.73   &  3.76  &  0.75   &  0.81   &  0.80   & 1.25  &  1.52  & 1.35  & 4.99  &  6.33  &  6.54  & 4.90 & 6.22 & 6.52 & 1.40 & 2.73  & 0.88   \\
Airport    & 372 & 372 & 411 & 2.73  &  3.27   &  3.89  &  0.74   &  0.86   &  0.85   & 1.28  &  1.75  & 1.49  & 4.44  &  5.34  &  6.28  & 4.41 & 5.28 & 6.22 & 1.39 & 3.27  & 0.83    \\
Babble     & 384 & 384 & 348 & 2.67  &  3.26   &  3.82  &  0.74   &  0.85   &  0.84   & 1.28  &  1.70  & 1.45  & 4.28  &  5.01  &  5.93  & 4.23 & 4.94 & 5.88 & 1.38 & 3.26  & 0.83  \\
Car        & 354 & 354 & 370 & 2.87  &  3.23   &  3.66  &  0.75   &  0.85   &  0.84   & 1.26  &  1.71  & 1.45  & 4.55  &  5.42  &  6.26  & 4.49 & 5.34 & 6.21 & 1.38 & 3.23  & 0.85    \\
Exhibition & 333 & 333 & 366 & 2.91  &  3.24   &  3.94  &  0.77   &  0.86   &  0.85   & 1.30  &  1.79  & 1.47  & 5.41  &  6.15  &  6.91  & 5.37 & 6.07 & 6.87 & 1.34 & 3.24  & 0.84   \\
SSN        & 386 & 386 & 359 & 2.60  &  3.14   &  3.20  &  0.73   &  0.83   &  0.82   & 1.23  &  1.64  & 1.39  & 3.15  &  4.31  &  4.81  & 3.05 & 4.18 & 4.65 & 1.47 & 3.14  & 0.97   \\
Station    & 380 & 380 & 349 & 2.64  &  3.19   &  3.78  &  0.74   &  0.85   &  0.85   & 1.25  &  1.68  & 1.49  & 4.08  &  5.09  &  6.39  & 4.01 & 4.99 & 6.33 & 1.39 & 3.19  & 0.83    \\
Street     & 380 & 380 & 356 & 2.81  &  3.18   &  3.63  &  0.75   &  0.84   &  0.83   & 1.26  &  1.72  & 1.41  & 4.67  &  5.66  &  6.11  & 4.63 & 5.59 & 6.09 & 1.37 & 3.17  & 0.84    \\
\bottomrule
\end{tabular}

\vspace{-1.3em}
\caption{Comparative noise type–wise analysis over three extension ranges of our proposed AP-BWE on the VCTK noisy dataset.}
\label{tab:Comparative_analysis_noise_source_wise-APBWE}
\vspace{-2.493em}
\end{table*}

\subsection{Details on Computational Complexity}
\label{append:analysisoncomputationcomplexity}

Table \ref{tab:computation} has the detailed analysis on computational complexity.

\textbf{Model Size (Parameters):} AP-BWE has 72.07M parameters, more than double that of CIS-BWE (33.74M). This means AP-BWE consumes $\sim$2× more memory for model weights and intermediate activations. CIS-BWE is more memory-efficient, making it easier to deploy on limited-resource devices.

\textbf{Computational Cost (MACs \& FLOPs):} AP-BWE requires 14.2B MACs and 28.5B FLOPs per inference, whereas CIS-BWE only requires 6.79B MACs and 13.6B FLOPs. Again, CIS-BWE achieves over 50\% reduction in operations, which directly reduces inference cost and memory bandwidth.

\textbf{Real-Time Factor (RTF, GPU):} AP-BWE runs at 0.0023–0.0025× RTF, while CIS-BWE achieves 0.0025–0.0028× RTF. Lower RTF means faster-than-real-time inference. Although both models are highly efficient on the GPU, CIS-BWE consistently shows a similar RTF compared to AP-BWE.  

\textbf{Inference time (GPU):} AP-BWE has an inference time of 14.93 ms and 16.6 ms, while CIS-BWE achieves 13.6 ms and 16.64 ms for 4-16 kHz and 16-48 kHz frequency range, respectively. Due to optimization of the generator and discriminators (see Sections \ref{subsec:generator_ablation}, \ref{subsec:Discriminator_Efficient}, \ref{subsec:Parameter Breakdown for Discriminators}, \ref{subsec:Parameter Breakdown for Generators}), CIS-BWE achieves performance gain in all eight subjective and objective metrics with a lower/similar inference time. 

\vspace{-0.650em}
\begin{table}[ht!]
\centering
\scriptsize
\setlength{\tabcolsep}{3pt}
\renewcommand{\arraystretch}{0.85}

\begin{tabular}{lcccccc}
\toprule
Freq. range &  LSD & STOI  & PESQ   & SI-SDR & SI-SNR & NISQA-MOS \\
\midrule
2-16 kHz & 1.1068 & 0.8739 & 1.66 & 8.1487 & 8.118 & 4.2979\\
2-48 kHz & 1.21 & 0.8526 & 1.1836 & 6.963 & 6.9662 & 3.9987\\
4-48 kHz & 1.099 & 0.933 & 1.4921 & 12.045 & 11.99 & 4.2294\\
8-48 kHz & 1.092 & 0.933 & 1.506 & 12.39 & 12.33 & 3.975\\
12-48 kHz & 0.873 & 0.9976 & 3.1253 & 17.085 & 16.95 & 4.4854\\
24-48 kHz & 0.6531 & 0.9989 & 4.1822 & 23.42 & 23.38 & 4.5254\\

\bottomrule
\end{tabular}

\vspace{-0.94em}
\caption{Performance over different frequency ranges.}
\label{tab:SoTA_different_fr_range}
\vspace{-02.84em}
\end{table}

\vspace{-0.3em}
\subsection{Study for Different Frequency Ranges} 
\label{subsec:Analysis for Different Frequency Ranges}
\vspace{-0.2em}

Table~\ref{tab:SoTA_different_fr_range} investigates performance across different frequency ranges, indicating a clear pattern: \emph{as the gap between the narrow and target band reduces, performance improves}. Therefore, the broadest reconstruction (2-48\,kHz) is the hardest, giving the poorest scores, and the narrowest reconstruction (24-48\,kHz) is the easiest, giving the best scores. 

\begin{table*}[ht]
\centering
\resizebox{\textwidth}{!}{
\begin{tabular}{l c l c c c c r}
\hline
\textbf{Discriminator} & \textbf{Stage} & \textbf{Layer Type} & \textbf{In→Out} & \textbf{Kernel} & \textbf{Stride} & \textbf{Padding} & \textbf{Params} \\
\hline
\multirow{15}{*}{MRLD (per scale)} 
  & Block 1 & Depthwise Conv1d    & 1→1   & 5   & 2 & 2 & 6      \\
  &         & Pointwise Conv1d    & 1→32  & 1   & 1 & 0 & 64     \\
  &         & BatchNorm1d + LReLU(0.1)        & 32→32 & –   & – & – & 64     \\
\cline{2-8}
  & Block 2 & Depthwise Conv1d    & 32→32 & 5   & 2 & 2 & 192    \\
  &         & Pointwise Conv1d    & 32→64 & 1   & 1 & 0 & 2\,112 \\
  &         & BatchNorm1d + LReLU(0.1)        & 64→64 & –   & – & – & 128    \\
\cline{2-8}
  & Block 3 & Depthwise Conv1d    & 64→64 & 5   & 2 & 2 & 384    \\
  &         & Pointwise Conv1d    & 64→128& 1   & 1 & 0 & 8\,320 \\
  &         & BatchNorm1d + LReLU(0.1)          &128→128& –   & – & – & 256    \\
\cline{2-8}
  & Block 4 & Depthwise Conv1d    &128→128& 5   & 2 & 2 & 768    \\
  &         & Pointwise Conv1d    &128→256& 1   & 1 & 0 &33\,024 \\
  &         & BatchNorm1d + LReLU(0.1)         &256→256& –   & – & – & 512    \\
\cline{2-8}
  & Final   & Depthwise Conv1d    &256→256& 3   & 1 & 1 &1\,024  \\
  &         & Pointwise Conv1d    &256→1  & 1   & 1 & 0 & 257    \\
  &         & BatchNorm1d     & 1→1   & –   & – & – & 2      \\
\cline{2-8}
\multicolumn{7}{r}{\textbf{MRLD total (per scale)}} & 47\,113 \\
\hline
\multirow{15}{*}{MSDFA (per scale)} 
  & Block 1 & Depthwise Conv2d    & 1→1   & 3×3 & 1 & 1 & 10     \\
  &         & Pointwise Conv2d    & 1→32  & 1×1 & 1 & 0 & 64     \\
  &         & BatchNorm2d + LReLU(0.2)           &32→32  & –   & – & – & 64     \\
\cline{2-8}
  & Block 2 & Depthwise Conv2d    &32→32  & 3×3 & 2 & 1 & 320    \\
  &         & Pointwise Conv2d    &32→64  & 1×1 & 1 & 0 &2\,112  \\
  &         & BatchNorm2d + LReLU(0.2)           &64→64  & –   & – & – & 128    \\
\cline{2-8}
  & Block 3 & Depthwise Conv2d    &64→64  & 3×3 & 2 & 1 & 640    \\
  &         & Pointwise Conv2d    &64→128 & 1×1 & 1 & 0 &8\,320  \\
  &         & BatchNorm2d + LReLU(0.2)            &128→128& –   & – & – & 256    \\
\cline{2-8}
  & Block 4 & Depthwise Conv2d    &128→128& 3×3 & 2 & 1 &1\,280  \\
  &         & Pointwise Conv2d    &128→256& 1×1 & 1 & 0 &33\,024 \\
  &         & BatchNorm2d + LReLU(0.2)           &256→256& –   & – & – & 512    \\
\cline{2-8}
  & Block 5 & Depthwise Conv2d    &256→256& 3×3 & 1 & 1 &2\,560  \\
  &         & Pointwise Conv2d    &256→1  & 1×1 & 1 & 0 & 257    \\
  &         & BatchNorm2d         & 1→1   & –   & – & – & 2      \\
\cline{2-8}
\multicolumn{7}{r}{\textbf{MSDFA total (per scale)}} & 49\,549 \\
\hline
\multirow{6}{*}{MRAD (per res)} 
  & Conv 1    & Conv2d  WeightNorm & 1→64  & 7×5 & 2×2 & 3×2 & 2\,304  \\
  & Conv 2    & Conv2d  WeightNorm & 64→64 & 5×3 & 2×1 & 2×1 &61\,504  \\
  & Conv 3    & Conv2d  WeightNorm & 64→64 & 5×3 & 2×2 & 2×1 &61\,504  \\
  & Conv 4    & Conv2d WeightNorm & 64→64 & 3×3 & 2×1 & 1×1 &36\,928  \\
  & Conv 5    & Conv2d  WeightNorm & 64→64 & 3×3 & 2×2 & 1×1 &36\,928  \\
  & Conv\_post& Conv2d  WeightNorm &64→1   & 3×3 & 1×1 & 1×1 &   577  \\
\cline{2-8}
\multicolumn{7}{r}{\textbf{MRAD total (per res)}} & 199\,745 \\
\hline
\multirow{6}{*}{MRPD (per res)} 
  & Conv 1    & Conv2d WeightNorm & 1→64  & 7×5 & 2×2 & 3×2 & 2\,304  \\
  & Conv 2    & Conv2d WeightNorm & 64→64 & 5×3 & 2×1 & 2×1 &61\,504  \\
  & Conv 3    & Conv2d WeightNorm & 64→64 & 5×3 & 2×2 & 2×1 &61\,504  \\
  & Conv 4    & Conv2d WeightNorm & 64→64 & 3×3 & 2×1 & 1×1 &36\,928  \\
  & Conv 5    & Conv2d WeightNorm & 64→64 & 3×3 & 2×2 & 1×1 &36\,928  \\
  & Conv\_post & Conv2d WeightNorm &64→1   & 3×3 & 1×1 & 1×1 &   577  \\
\cline{2-8}
\multicolumn{7}{r}{\textbf{MRPD total (per res)}} & 199\,745 \\
\hline
\multicolumn{7}{r}{\textbf{Grand total (all discriminators)}} & 1\,681\,780 \\
\hline
\end{tabular}
}

\caption{Layer-wise parameter breakdown, per-discriminator totals, and grand total for all four discriminators in CIS-BWE.}
\label{tabel:layerwisediscriminatorbreakdown}
\end{table*}

\begin{table*}[t]
\centering
\resizebox{\textwidth}{!}{
\begin{tabular}{l l c c c c c r}
\hline
\textbf{Stage / Component}         & \textbf{Layer Type}               & \textbf{In→Out} & \textbf{Kernel} & \textbf{Stride} & \textbf{Padding} & \textbf{Heads} & \textbf{Params} \\
\hline
\multicolumn{8}{c}{\textbf{Pre‐processing}} \\
\hline
Pre‐mag convolution                & Conv1d                            & 513→512         & 7×7             & 1×1             & 7×1             & –              & 1\,839\,104     \\
Pre‐pha convolution                & Conv1d                            & 513→512         & 7×7             & 1×1             & 7×1             & –              & 1\,839\,104     \\
Pre‐mag LayerNorm                  & LayerNorm                         & 512→512         & –               & –               & –               & –              &     1\,024      \\
Pre‐pha LayerNorm                  & LayerNorm                         & 512→512         & –               & –               & –               & –              &     1\,024      \\
\hline
\multicolumn{8}{c}{\textbf{ConformerNeXtBlock (per block breakdown)}} \\
\hline
\multicolumn{8}{c}{\textbf{FFN \& Norm}} \\
$\text{Norm}_1$                             & LayerNorm                         & 512→512         & –               & –               & –               & –              &     1\,024      \\
$\text{FFN}_1$ - $\text{Linear}_1$ + \text{GELU} + \text{Droput(0.1)}                     & Linear                            & 512→2\,048      & –               & –               & –               & –              & 1\,050\,624     \\
$\text{FFN}_1$ - $\text{Linear}_2$ + \text{Dropout(0.1)}                     & Linear                            & 2\,048→512      & –               & –               & –               & –              & 1\,049\,088     \\
$\text{Norm}_2$                             & LayerNorm                         & 512→512         & –               & –               & –               & –              &     1\,024      \\
\hline
\multicolumn{8}{c}{\textbf{Self‐Attention}} \\
Self‐Attention                     & MultiHeadSelfAttention (embed=512)& 512→512         & –               & –               & –               & 8              & 1\,050\,624     \\
\hline
\multicolumn{8}{c}{\textbf{ConvNeXt Components}} \\
Depthwise Conv1d                   & Depthwise Conv1d                  & 512→512         & 7               & 1               & 3               & –              &     4\,096      \\
ConvNeXt—Norm                      & LayerNorm                         & 512→512         & –               & –               & –               & –              &     1\,024      \\
ConvNeXt—PWConv1 + GELU            & Linear                            & 512→1\,536      & 1               & 1               & 0               & –              &   787\,968      \\
ConvNeXt—PWConv2                   & Linear                            & 1\,536→512      & 1               & 1               & 0               & –              &   786\,944      \\
ConvNeXt—Gamma                     & Learned scale                     & 512→512         & –               & –               & –               & –              &       512       \\
\hline
\multicolumn{7}{r}{\textbf{Total per ConformerNeXtBlock}}                                    & 6\,834\,688     \\
\multicolumn{7}{r}{\textbf{ConformerNeXtBlock total (4 blocks)}}                           &27\,338\,752     \\
\hline
\multicolumn{8}{c}{\textbf{LatticeBlock1D}} \\
\hline
LatticeBlock1D (per block)         & Two‐branch fusion using one        & 512→512         & –               & –               & –               & –              & $4\times4$    \\
                                   & ConformerNeXt Block + 4 scalars     &                 &                 &                 &                 &                &                 \\
\cline{7-8}
\multicolumn{7}{r}{\textbf{Total LatticeBlock1D (4 blocks)}}                               &27\,338\,768     \\
\hline
\multicolumn{8}{c}{\textbf{Post‐processing}} \\
\hline
Post‐mag LayerNorm                & LayerNorm                         & 512→512         & –               & –               & –               & –              &     1\,024      \\
Post‐mag FFN                      & Linear                            & 512→513         & –               & –               & –               & –              &   263\,169      \\
Post‐pha LayerNorm                & LayerNorm                         & 512→512         & –               & –               & –               & –              &     1\,024      \\
FFN\_r post-pha (real)            & Linear                            & 512→513         & –               & –               & –               & –              &   263\,169      \\
FFN\_i post‐pha (imag)            & Linear                            & 512→513         & –               & –               & –               & –              &   263\,169      \\
\hline
\multicolumn{7}{r}{\textbf{Total generator parameters}}                                   &31\,808\,531     \\
\hline 
\end{tabular}} 
\caption{Layer‐wise parameter breakdown for the CIS-BWE generator, including LatticeBlock1D parameters alongside Pre‐processing, ConformerNeXt blocks, and Post‐processing.}

\label{table:generatorbraekdown}
\end{table*}

\newcolumntype{L}{>{\raggedright\arraybackslash}X}
\newcolumntype{C}{>{\centering\arraybackslash}X}

\begin{table*}[ht]
  \centering
  \small
  \begin{tabularx}{\textwidth}{L C L}
    \toprule
    \textbf{Hyperparameter} & \textbf{Value} & \textbf{Use Case \& Rationale} \\
    \midrule
    Number of GPUs
      & 1
      & Ensure faster training and inference by effectively leveging parallel processing of CUDA cores \\
    Max epochs
      & 50
      & Provide enough weight updates for convergence yet avoid overfitting. \\
    Batch size
      & 16
      & Balance between gradient stability with computational resource constraints. \\
    Initial learning rate
      & $2\times10^{-4}$
      & Find balance between convergence with training stability. \\
    Adam $\beta_{1}$
      & 0.8
      & Optimizer momentum parameter set to adapt quickly to adversarial non-stationarity. \\
    Adam $\beta_{2}$
      & 0.99
      & Optimizer second moment estimate parameter for stable variance control. \\
    Learning–rate decay
      & 0.999
      & Decrease LR to subtly fine-tune weights toward convergence. \\
    Random seed
      & 1234
      & Ensure to generate same results across different run \\
    ConvNeXt channels
      & 512
      & Provide enough capacity to capture features. \\
    ConformerNeXt blocks
      & 4
      & Provide enough parameter without sacrificing performance.  \\
    Segment size (samples)
      & 8000
      & Capture sufficient audio context for effective BWE.\\
    FFT size (n\_fft)
      & 1024
      & Balance between frequency resolution against computational load. \\
    Hop length
      & 80
      & Overlap chosen to smooth ripple effects without increasing computational load. \\
    Window length
      & 320
      & Balance time–frequency resolution in STFT. \\
    High-rate sampling rate (Hz)
      & 16 K / 48 K
      & Define wide-band frequency ranges. \\
    Low-rate sampling rate (Hz)
      & 2 K / 4 K / 8 K / 16 K / 24 K
      & Different differncy range to evaluate the robustness of the model. \\
    Subsampling ratio
      & 2 / 4 / 8 / 12 / 24
      & Downsampling factors corresponding to low-rate configurations. \\
    Number of data-loading workers
      & 4
      & Parallel I/O to maximize throughput \\
    Distributed backend
      & \texttt{nccl}
      & Efficient GPU-to-GPU communication\\
    Distributed init URL
      & \texttt{tcp://localhost:54321}
      & Local rendezvous for single-node distributed setup. \\
    Distributed world size
      & 1
      & Single-process distributed for clean scaling. \\
    MRLD window sizes
      & 64, 128, 256, 512, 1024
      & Multi-scale Lyapunov analysis to capture deterministic chaotic features at different resolutions. \\
    MSDFA scales
      & 100, 200, 300, 500, 600
      & Range of DFA scales for fractal dimension analysis in discriminator. \\
    MRAD resolutions (n\_fft, hop, win)
      & (512,128,512), (1024,256,1024), (2048,512,2048)
      & Multi-resolution STFT settings to capture amplitude dynamics. \\
    MRPD resolutions (n\_fft, hop, win)
      & (512,128,512), (1024,256,1024), (2048,512,2048)
      & Multi-resolution STFT settings to capture phase dynamics. \\
    \bottomrule
  \end{tabularx}
  
  \caption{Training and model hyperparameters for the CIS-BWE setup, with use cases and rationale}
  \label{tab:hyperparams}
\end{table*}

\subsection{Dataset Class and DataLoader}
\label{sec:dataset_dataloader}

Above preprocessing steps are defined in a custom PyTorch \texttt{Dataset} class. After initialization, we shuffle audio file lists with a fixed random seed (\texttt{random.seed(1234)}) for ensuring reproducibility. The \texttt{\_\_getitem\_\_} function handles loading of data (or reuse of cache), apply resampling, apply segmentation, and convert to mono channel, and return a tuple of 1-D tensors, which contains 8{,}000 samples. Total length of the dataset (\texttt{\_\_len\_\_}) is equal to the number of files in each split. During training process, we initialize a \texttt{DataLoader} class which uses four worker processes (\texttt{num\_workers=4}) and uses \texttt{DistributedSampler} to ensure distribution of distinct partitions of the dataset.

\subsection{Computational Analysis only for LE and DF Features in MRLD and MSDFA}
\label{append:analysisoncomputationcomplexityepochtime}

{\color{black}

We devise new experiments and found out the overhead of the chaos-informed feature extraction using a dedicated python profiling script over the full test portion of the VCTK dataset (2{,}937 files; total duration $\approx$ 6{,}548.8~s). We report both per-segment size of 8000 samples and full-dataset statistics, as well as real-time factors (RTF).

\paragraph{(a) Profiling setup.}
\begin{itemize}
  \item Sampling rate: 16~kHz.
  \item Segment size used by MRLD and MSDFA: 8{,}000~samples.
  \item Hardware: NVIDIA RTX-4090 GPU and AMD Ryzen 9 7950X3D 16-Core CPU.
  \item All chaotic features are computed \emph{on-the-fly during training}, on the GPU. The inference-time reported in the manuscript do not correlate with the results reported here, as during inference we only use the generator part of the CIS-BWE, but during training we use chaotic features in discriminators to guide the generator to capture long-range dependencies and reducing over-smoothed spectra.
  \item Finally, we note that most of the recent literature reports FLOPs and MACs using the
\verb|get_model_complexity_info| function from the \texttt{ptflops} Python library. This tool is
designed for parameterized neural-network modules, and it returns zero FLOPs/MACs when applied to
our Lyapunov Exponent (LE) feature block because that computation is parameter-free. Therefore, for
LE we instead report an analytic approximation based on the underlying operations.

Specifically, for an input of length $T$ (samples) and batch size $B$, with window sizes
$\mathcal{W} = \{w_s\}$, embedding dimension $d$ and time delay $\tau$, we define
\begin{align*}
  n_w(w_s) &= \left\lfloor \frac{T}{w_s} \right\rfloor,\\
  M(w_s)   &= w_s - (d-1)\tau ,
\end{align*}
where $n_w(w_s)$ is the number of windows of length $w_s$ that fit into the signal and $M(w_s)$ is
the number of valid embedded time steps per window. The main cost per window is dominated by the
pairwise distance computation and divergence loops, both scaling as $\mathcal{O}(M^2 d)$. We
approximate the total MAC count for the Lyapunov feature computation as:

\begin{equation}
\small
  \mathrm{MACs}_{\mathrm{LE}}(T)
  \;\approx\;
  \alpha\, B \sum_{w_s \in \mathcal{W}}
  n_w(w_s)\,\bigl(M(w_s)\bigr)^{2}\, d ,
\end{equation}

where $\alpha \approx 5$ is a small heuristic constant that absorbs lower-order operations. The corresponding FLOP estimate is:

\begin{equation}
\small
  \mathrm{FLOPs}_{\mathrm{LE}}(T)
  \;\approx\; 2\,\mathrm{MACs}_{\mathrm{LE}}(T).
\end{equation}

For a single segment of length $T$ with batch size $B=1$, this simplifies to:

\begin{equation}
\small
  \mathrm{MACs}_{\mathrm{LE,seg}}
  \;\approx\;
  \alpha \sum_{w_s \in \mathcal{W}}
  \left\lfloor \frac{T}{w_s} \right\rfloor
  \left(w_s - (d-1)\tau\right)^{2} d ,
\end{equation}

\begin{equation}
\small
  \mathrm{FLOPs}_{\mathrm{LE,seg}}
  \;\approx\; 2\,\mathrm{MACs}_{\mathrm{LE,seg}}.
\end{equation}

\end{itemize}

\paragraph{(b) Per-segment computational cost.}
From the measured Multiply--Accumulate operations (MACs) in our profiling logs:

\begin{itemize}
  \item \textbf{LE}
  \begin{itemize}
    \item Per-segment MACs: $\approx 707.7$ million.
    \item Per-segment FLOPs (approx.\ $2\times$MACs): $\approx 1415.5$ million.
    \item Approx.\ GPU latency time per segment: 6.56~ms.
  \end{itemize}

  \item \textbf{DFA}
  \begin{itemize}
    \item Per-segment MACs: $\approx 0.277$ million.
    \item Per-segment FLOPs: $\approx 0.554$ million.
    \item Approx.\ GPU latency time per segment: 0.145~ms.
  \end{itemize}
\end{itemize}

Thus, LE is the dominant contributor to the additional chaotic overhead; DFA is negligible in comparison.
These operations are used \emph{only} inside the discriminators during training, so they do not affect runtime deployment. They do, however, explain the observed difference in training time between AP-BWE and CIS-BWE:
AP-BWE trains at $\approx 17$~min/epoch, while CIS-BWE (with MRLD+MSDFA) trains at $\approx 25$~min/epoch.

\begin{itemize}
  \item Training-time overhead is primarily due to LE estimation in MRLD (MSDFA overhead is negligible).
  \item At inference, CIS-BWE uses $\approx 0.5\times$ fewer parameters, MACs, and FLOPs than AP-BWE, and no chaotic operations, hence lower memory and comparable or better RTF which makes CIS-BWE better candidate for real-world deployment in edge devices.
\end{itemize}

}

\subsection{Time–Frequency Feature Extraction and Reconstruction}
\label{sec:feature_extraction}

We design a feature extraction function, \texttt{amp\_pha\_stft}, which calculates the short‐time Fourier transform (STFT) on audio segments using [win\_size, hop\_size, fft] = [320, 80, 1024] and a Hann window parameter. Using generated complex spectrogram \(X \in \mathbb{C}^{F \times T}\), we derive two features for giving input to dual stream CIS-BWE.
The is log‐amplitude calculate using 
\[{M\_nb} = \log\bigl(\lvert X\rvert + 10^{-4}\bigr)\]
and instantaneous phase
\[{\Phi\_nb} = \arg(X).\]

The output of the CIS-BWE are converted back into audio waveform by \texttt{amp\_pha\_istft}, which exponentiates the predicted log‐amplitude, reconstructs the complex spectrogram, and applies an inverse STFT using the same windowing parameters to obtain the final HR audio.

\subsection{Inference Workflow}
\label{sec:inference_workflow}

We load the trained CIS-BWE checkpoint to generate wide-band audio using narrow-band \texttt{.wav} inputs. The script initially loads the trained \texttt{CIS\_BWE Model} checkpoint onto the specified device (GPU or CPU). It then recursively searches the input directory for narrowband \texttt{.wav} files. For each discovered file, the script:

\begin{enumerate}
  \item Applies similar pre-processing and resampling steps for HR and LR creation.
  \item Extracts log-amplitude and phase spectrograms as features via \texttt{amp\_pha\_stft}.
  \item Feeds the two spectrogram features into each stream of the generator network.
  \item Maps the narrowband audios to wideband by generating missing high frequency components
  \item Applies \texttt{amp\_pha\_istft} to invert the output representations to waveforms.
  \item Saves the output audio as 16 bit PCM \texttt{.wav} files at 16/48 KHz.
  \item Logs the losses and total processing time for extensive analysis later.
\end{enumerate}

\subsection{Parameter Breakdown for Discriminators}
\label{subsec:Parameter Breakdown for Discriminators}

A layerwise parameter breakdown for each discriminator and grand total for all four discriminators in CIS-BWE are shown in Table \ref{tabel:layerwisediscriminatorbreakdown}.

\subsection{Parameter Breakdown for Generators}
\label{subsec:Parameter Breakdown for Generators}

A layer-wise parameter breakdown for the CIS-BWE generator, including LatticeBlock1D parameters alongside pre-processing, ConformerNeXt blocks, and post-processing are shown in Table  \ref{table:generatorbraekdown}.

\subsection{Hyperparameters and Configuration}
\label{subsec:hyperparameters}

\textbf{Software Version:} We use an Anaconda virtual environment with Python 3.9.21, PyTorch 2.0.0+cu118, Torchaudio 0.15.0+cu118, Torchvision 0.15.0+cu118, and CUDA Toolkit 11.8.0.

For distributed training and potential scalability, we use the NCCL for multi-GPU training and TCP to initialize communication between processes.

The training and model hyperparameters for the CIS-BWE setup, with use cases and rationale are provided in Table \ref{tab:hyperparams}.

\subsection{Subjective evaluation details} 
\label{subsec:subj_eval_details}
\begin{figure*}[htbp]
  \centering
\includegraphics[width=0.98\textwidth,height=0.58\textheight]{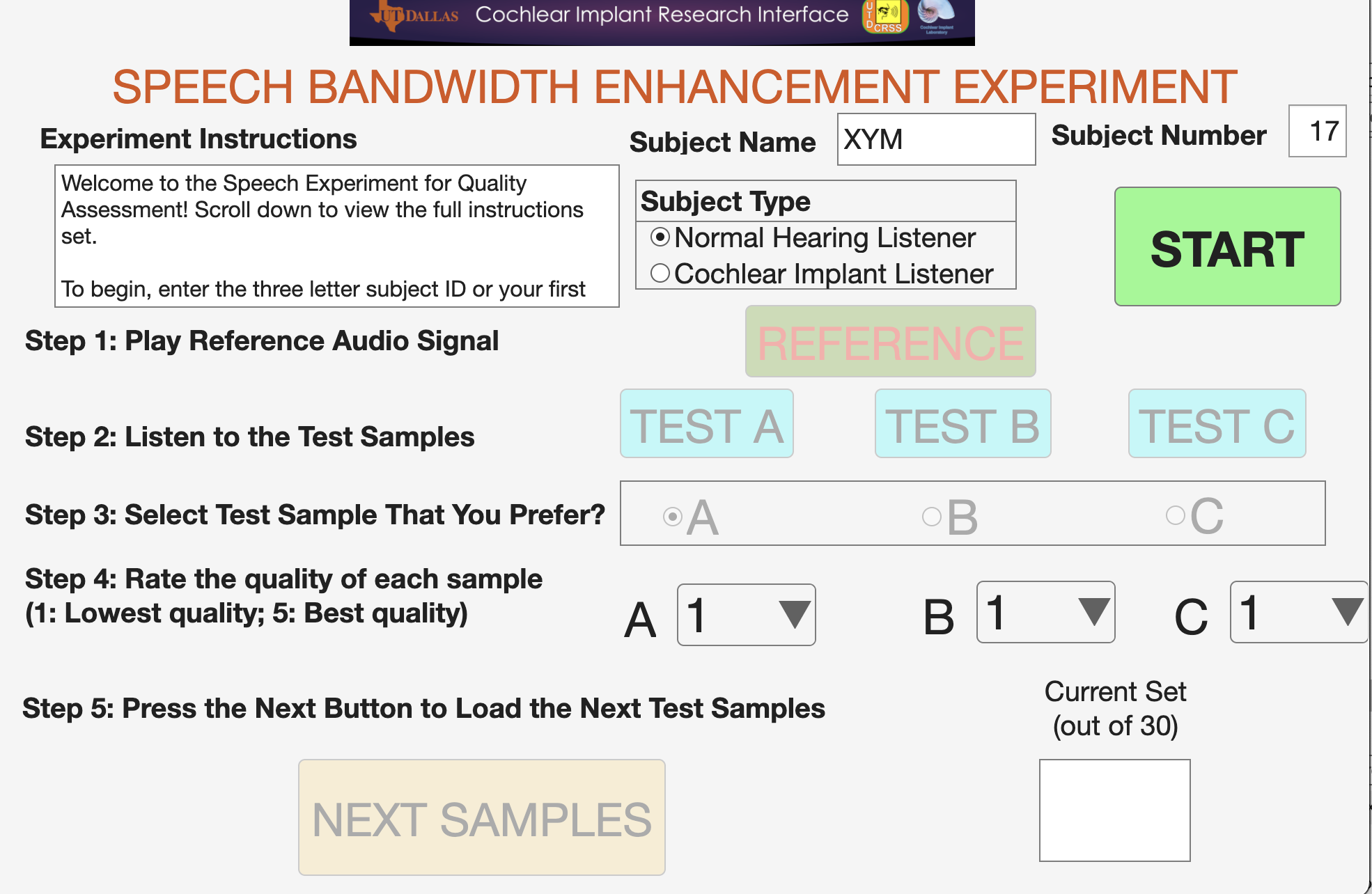}
\vspace{-0.750em}
  \caption{The MATLAB interface used in Subjective Tests.}
  \label{fig:subj_eval_interface_ss}
  \vspace{-01.60em}
\end{figure*}

To evaluate the performance of the proposed CIS-BWE, a formal pair-preference listening test was conducted. A total of ten Bangladeshi under graduate student who self-reported normal-hearing (NH) participants—comprising four males and six females with an average age of 24—participated in the study. The participants voluntarily join to rate the audios without any compensation. At first they are trained on how to assign scores based on perceived perceptual quality of audios. They are also briefed about the purpose of the experiments, potential risks, and about the outcome of this paper.
All participants were non-native English speakers and used soundproof headsets to ensure consistent and distraction-free listening conditions.
Each participant evaluated 30 sets of speech samples. Every set included three randomly presented versions of the same utterance: (i) the unprocessed (noisy) signal, (ii) the baseline-enhanced signal using APBWE processing, and (iii) the speech processed by the proposed network. For reference, a clean version of the speech signal was also available, though it was not part of the evaluation. Participants rated the perceptual quality of each sample on a 5-point Mean Opinion Score (MOS) scale, where 1 indicates the lowest and 5 the highest quality. Additionally, they were asked to select the most preferred version from the three presented options.
The test used 5–7-second speech clips selected from the VCTK dataset, which were processed under three different frequency band conditions: 2–16 kHz, 12–48 kHz, and 24–48 kHz. Individual pair-preference results were analyzed separately for both male and female participants across all band configurations. In the figure \ref{fig:subj_eval_interface_ss}, we have presented the MATLAB interface that we use to conduct the subjective evaluation. 

The findings in figure \ref{fig:comparison} clearly show that the speech enhanced by the proposed network was consistently and significantly preferred over both the unprocessed and baseline-processed versions. In particular, the proposed CIS-BWE achieved a 54.5\% improvement in user preference compared to unprocessed speech and a 2\% improvement over the APBWE baseline. These results highlight the network’s robust ability to enhance perceptual speech quality under noisy and reverberant conditions.

\end{document}